\documentclass[10pt]{article}
\usepackage{latexsym}
\newtheorem{thm}{Theorem}
\newtheorem{cor}{Corollary}
\newtheorem{lem}{Lemma}
\newtheorem{prop}{Proposition}
\newtheorem{defn}{Definition}
\newtheorem{rem}{Remark}
\newtheorem{example}{Example}

\newcommand{\Real}{\mathbf R}
\newcommand{\To}{\longrightarrow}
\newcommand{\ii}{\imath}

\newcommand{\smooth}[1][]{\ensuremath{\textrm{C}^{\infty}{#1}\;}}
\newcommand{\tang}[2][]{\ensuremath{\textrm{T}_{#1}(#2)}}
\newcommand{\ctang}[2][]{\ensuremath{\textrm{T}^*_{#1}(#2)}}
\newcommand{\gr}[2]{\ensuremath{\textrm{Gr}_{#1}(#2)}}
\newcommand{\immerse}{\hookrightarrow}
\newcommand{\scalar}[2]{\langle\,{#1}\,,\,{#2}\,\rangle}
\newcommand{\der}[1][]{\textrm{d}{#1}}
\newcommand{\ud}{\textrm{d}}
\newcommand{\then}{\quad\Rightarrow\quad}

\newcommand{\vecf}[1]{V^1(#1)}
\newcommand{\vecfn}[1]{V_N^1(#1)}
\newcommand{\same}{\Longleftrightarrow}
\newcommand{\lie}[1]{\mathit{#1}}
\newcommand{\lief}{\Omega^1(L,A)}
\newcommand{\sublief}{\Omega^1(L,A,L_0)}
\newcommand{\sublieff}{\sublief{/}I{\cdot}\sublief}
\newcommand{\formsD}{\Omega^1_D(A)}
\newcommand{\formsI}{\Omega^1(A,I)}
\newcommand{\cd}[1][]{\ensuremath{\partial_{#1}}}
\newcommand{\op}[1]{\ensuremath{\widehat{#1}}}
\newcommand{\mend}[2][]{\ensuremath{\textrm{End}_{#1}{(#2)}}}
\newcommand{\diff}[1]{\textrm{Diff}^1(#1)}
\newcommand{\iso}{\cong}
\newcommand{\subforms}[2][]{\widehat{\Omega}^{#1}(#2)}
\newcommand{\alt}[2][]{\textrm{Alt}^{#1}(#2)}
\newcommand{\xalt}[2][]{\textrm{Alt}^{#1}_A(#2)}
\newcommand{\bop}[2]{\widetilde{#1}(#2)}
\newcommand{\rk}[1]{\ensuremath{\textrm{Rank}(#1)}}
\newcommand{\dist}[2][]{\ensuremath{\mathcal{D}_{#1}(#2)}}
\newcommand{\comp}[1]{\ensuremath{C_0^{\infty}(#1)}}
\newcommand{\vol}{\textrm{vol}}
\newcommand{\poiss}[2]{\{\,#1\,,\,#2\,\}}
\newcommand{\genf}{generalized function}

\newcommand{\mandim}[1]{\textbf{dim}(#1)}

\begin{document}
\title{Noncommutative Geometry of Phase Space}
\author{Zakaria Giunashvili}
\date{\today}
\maketitle
\begin{abstract}
We investigate the geometric, algebraic and homologic structures
related with Poisson structure on a smooth manifold. Introduce a
noncommutative foundations of these structures for a Poisson
algebra. Introduce and investigate \textbf{noncommutative Bott
connection} on a foliated manifold using the algebraic definition
of submanifold and quotient manifold. Develop an algebraic
construction for the reduction of a degenerated Poisson algebra.
\end{abstract}
\newpage
\tableofcontents
\newpage
\section{Distributions on \smooth- Class Manifolds: General Overview}
In this section we give a brief overview of some definitions and
facts concerning the distributions on \smooth- class manifold. We
consider not only regular distributions (i.e., the distributions
with a constant rank), but singular distributions too (i.e., the
distributions the rank of which varies from point to point).

For any vector space $V$ and $k \in \mathbf{N}$, we denote by
$\gr{k}{V}$ the \emph{Grassmann manifold} of $k$-dimensional
vector subspaces of the vector space $V$.
\begin{defn}
Let $M$ be a \smooth- class manifold and $\pi:E\To M$ be a vector
fiber bundle. The fiber bundle $\widetilde{\pi}_k:\gr{k}{E}\To M$
is called the \emph{Grassmanization} of the fiber bundle
$\pi:E\To~M$, or the \emph{Grassmanian fiber bundle} corresponding
to $\pi:E\To~M$, if for each point $x_0\in~M$ its fiber
$\widetilde{\pi}^{-1}(x_0)$ is the Grassmann manifold of
$k$-dimensional subspaces of the vector space $\pi^{-1}(x_0)$.
\end{defn}
Let $\tau:\tang{M}\To M$ be the tangent vector bundle over the
manifold $M$.
\begin{defn}
For an integer number $k$ such that $1\leq~k~\leq~n$, where
$n=\dim(M)$, a $k$-dimensional distribution $D$, on the manifold
$M$, is a correspondence $x\mapsto~D(x)$, where $x~\in~M$ and
$D(x)$ is a $k$-dimensional subspace of the tangent space
$\tang[x]{M}$.\\
Consider the Grassmanization of the tangent bundle over the
manifold $M$
$$
\widetilde{\tau}_k:\gr{k}{\tang{M}}\To M
$$
A $k$-dimensional distribution $D$, can be considered as a section
of this fiber bundle:
$$
M\ni x~\mapsto~D(x)\in\gr{k}{\tang[x]{M}}
$$
The distribution $D$ is said to be a \smooth- class, or
\emph{smooth} distribution, if $D$ is a \smooth- class section.
\end{defn}
\begin{defn}
A smooth distribution $D$ is called \textbf{involutive}, if for any
two smooth vector fields $X$ and $Y$ on the manifold $M$, such
that for each $x\in M$ the vectors $X(x)$ and $Y(x)$ are elements
of the vector space $D(x)$, their commutator $[X,Y]$ is also such
that for each $x\in M$ the vector $[X,Y](x)$ is an element of the
vector space $D(x)$.
\end{defn}
\begin{defn}
A submanifold $N$ of the manifold $M$ is said to be an
\emph{integral} submanifold for a given distribution $D$ on $M$ if
for each point $x\in N$, we have that $\tang[x]{N}=D(x)$.\\
A distribution $D$ on the manifold $M$ is said to be
\emph{integrable} if for each point $x\in M$ there exists an
integral submanifold $N$ for the distribution $D$, such that $x\in
N$.
\end{defn}
The Frobenius' classical theorem states that, a \smooth- class
distribution $D$ on a \smooth- class manifold is integrable if and
only if $D$ is involutive. Moreover, if the distribution $D$ is
involutive, then for each point $x_0\in M$, there exists its
neighborhood $U$ and a coordinate system
\begin{displaymath}
    u_1,u_2,\ldots,u_n\;:\;U\To\Real
\end{displaymath}
such that the level submanifolds
\begin{displaymath}
    u_i=\mathbf{const},\quad i=k+1,\ldots,n
\end{displaymath}
are the integral submanifolds of the distribution $D$. Moreover:
if $N$ is a connected integral submanifold, such that $N\subset
U$, then $N$ is inside of one of these level submanifolds.

E. Cartan's formalism gives a different approach to the local
properties of distributions. This formalism is more general, and
is designed for studying the geometric properties of not only
distributions, but also higher-order differential equations (see,
for example \cite{hermann1} and \cite{kuranishi}).
\begin{defn}[Cartan distribution]
\label{cartan_distrib} For any point
$V\in\gr{k}{\tang{M}}$, let
$$K_V=\big((\widetilde{\tau}_k)'(V)\big)^{-1}(V)$$
be the subspace of the vector space $\tang[V]{\gr{k}{\tang{M}}}$
where
\begin{displaymath}
    (\widetilde{\tau}_k)'(V)\,:\,\tang[V]{\gr{K}{\tang{M}}}\To\tang[x]{M}
\end{displaymath}
is the differential of the projection mapping
\begin{displaymath}
\widetilde{\tau}_k\,:\,\gr{k}{\tang{M}}\To~M
\end{displaymath}
at the point $V$, and $x=\widetilde{\tau}_k(V)$.\\
The distribution
\begin{displaymath}
\gr{k}{\tang{M}}\ni
V\,\mapsto\,K_V\subset\tang[V]{\gr{k}{\tang{M}}}
\end{displaymath}
is called the \textbf{Cartan distribution} on the Graasmanization of
the tangent bundle $\tang{M}$.
\end{defn}
As the fiber of the fiber bundle
$\widetilde{\tau}_k:\gr{k}{\tang{M}}\To M$ is the Grassmann
manifold with dimension equal to $k(n-k)$, the dimension of the
total space $\gr{k}{\tang{M}}$ is equal to $n+k(n-k)$
\begin{lem}
For each $W\in\gr{k}{\tang{M}}$ the dimension of the subspace
$K_W$, correspondent to the Cartan distribution at the point $W$
is equal to \mbox{$k(n-k+1)$}.
\end{lem}
\textbf{Proof.} Let $V_W$ be the vertical tangent space of the Grassmannian
fiber bundle
\begin{displaymath}
    \widetilde{\tau}_k\,:\,\gr{k}{\tang{M}}\To M
\end{displaymath}
at the point $W\in\gr{k}{\tang[x]{M}}$. We have the following
exact sequence
\begin{equation}\label{exact_sequence1}
0\;\To\;V_W\;\stackrel{\ii}{\immerse}\;K_W\;\stackrel{(\tilde{\tau}_k)'(W)}{\To}\;W\;\To\;0
\end{equation}
from which follows that $\dim(K_W)=\dim(V_W)+\dim(W)$. The space
$V_W$ is the tangent space of the Grassmann manifold
$\gr{k}{\tang[x]{M}}$ at the point $W$, and as it is well-known,
is isomorphic to the space $Hom\big(W,\,\tang[x]{M}/W\big)$.
Therefore, we have that
\begin{displaymath}
\dim\big(K_W\big)=\dim\big(Hom\big(W,\tang[x]{M}/W\big)\big)+\dim\big(W\big)=k(n-k+1)
\end{displaymath}
$\Box$

Let $X$, $Y$ and $Z$ be finite-dimensional real vector spaces and
the following is an exact sequence of linear mappings
\begin{equation}\label{exact_sequence2}
0\;\To\;Hom\big(Z,\,X\big)\;\stackrel{\ii}{\immerse}\;Y\;\stackrel{\pi}{\To}\;Z\;\To\;0
\end{equation}
Any \textbf{splitting}, $s:Z\To Y$, of this exact sequence defines
an antisymmetric, bilinear form $\scalar{\cdot}{\cdot}_s$ on the
vector space $Y$, which takes its values in the vector space $X$:
\begin{equation}\label{symplectic_form1}
    \scalar{x}{y}_s=\big(x-(s\pi)(x)\big)\big(\pi(y)\big)-\big(y-(s\pi)(y)\big)\big(\pi(x)\big)
\end{equation}
for any $x,y\in Y$.

A subspace $L\subset Y$ is called \emph{isotropic} for the
bilinear form $\scalar{\cdot}{\cdot}_s$, if
$\scalar{y_1}{y_2}_s=0$ for every pair $(y_1,y_2)\in Y\times Y$.
\begin{lem} \label{lemma_splitting_independence}
If $s_1:Z\To Y$ and $s_2:Z\To Y$ are two splittings of the exact
sequence (\ref{exact_sequence2}), such that the subspace
$\mathbf{Image}(s_1)$ is isotropic for the bilinear form
$\scalar{\cdot}{\cdot}_{s_2}$, then the subspace
$\mathbf{Image}(s_2)$ is isotropic for the bilinear form
$\scalar{\cdot}{\cdot}_{s_1}$ and the two bilinear forms
$\scalar{\cdot}{\cdot}_{s_2}$ and $\scalar{\cdot}{\cdot}_{s_1}$,
coincide.
\end{lem}
\textbf{Proof.} Any splitting $s:Z\To Y$ defines an isomorphism
\begin{displaymath}
Y\,\,\cong\,\,Z\times\,Hom\big(Z,\,X\big)
\end{displaymath}
and the bilinear form $\scalar{\cdot}{\cdot}_s$ on the space $Y$
is the pull-back of the bilinear form $\scalar{\cdot}{\cdot}$ on
the space $Z\times Hom\big(Z,\,X\big)$, which is defined as
\begin{displaymath}
\scalar{(u,\alpha)}{(v,\beta)}=\alpha(v)-\beta(u)
\end{displaymath}
Any subspace $Z'\,\subset\,Z\times Hom\big(Z,\,X\big)$, which is a
complement of the subspace $\{0\}\,\times\,Hom\big(Z,\,X\big)$ can
be given as
\begin{displaymath}
Z'=\big\{(z,f(z))\;|\;z\in Z\big\}
\end{displaymath}
where $f:Z\To Hom(Z,\,X)$ is a linear mapping.\\
For any two elements $\big(z_1,f(z_1)\big)$ and
$\big(z_2,f(z_2)\big)$ from the subspace $Z'$ we have
\begin{displaymath}
\scalar{\big(z_1,f(z_1)\big)}{\big(z_2,f(z_2)\big)}=f\big(z_1\big)(z_2)-f\big(z_2\big)(z_1)
\end{displaymath}
Therefore, the subspace $Z'$ is isotropic if and only if
\begin{displaymath}
f\big(z_1\big)(z_2)=f\big(z_2\big)(z_1)
\end{displaymath}
for every $z_1$ and $z_2$ from the space $Z'$.

Any element $(z,\alpha)\in Z\times Hom\big(Z,\,X\big)$ can be
represented as
\begin{displaymath}
\big(z,f(z)\big)+\big(0,\alpha-f(z)\big)
\end{displaymath}
where $\big(z,f(z)\big)\in Z'$ and $\big(0,\alpha-f(z)\big)\in
Hom\big(Z,\,X\big)$. Therefore, the bilinear form defined by the
subspace $Z'$ is
\begin{eqnarray*}
\lefteqn{\scalar{(z_1,\alpha)}{(z_2,\beta)}_{Z'}=\big(\alpha-f(z_1)\big)(z_2)-\big(\beta-f(z_2)\big)(z_1)=}\\
&&=\scalar{(z_1,\alpha)}{(z_2,\beta)}-\big(f\big(z_1\big)(z_2)-f\big(z_2\big)(z_1)\big).
\end{eqnarray*}
which shows that, the bilinear forms $\scalar{\cdot}{\cdot}_{Z'}$
and $\scalar{\cdot}{\cdot}$ are equal if and only if the subspace
$Z'$ is an isotropic subspace for the bilinear form
$\scalar{\cdot}{\cdot}$.
$\Box$

Let us introduce the following fiber bundles over the total space
of the Grassmanization \gr{k}{\tang{M}}:\\
$\pi_\Theta:\Theta\To\gr{k}{\tang{M}}$ be the
\emph{canonical} fiber bundle over the Grassmannization of the
tangent bundle $\tang{M}$. That is: the fiber of the bundle
$\pi_\Theta$ at a point $W\in\gr{k}{\tang[x]{M}}$ is the vector space $W$;\\
$\pi_K:K\To\gr{k}{\tang{M}}$ be the fiber bundle corresponding to
the Cartan distribution (see Definition~\ref{cartan_distrib}) on
the manifold $\gr{k}{\tang{M}}$. That is: the fiber at a point
$W\in\gr{k}{\tang[x]{M}}$ is the Cartan subspace $K_W$;\\
$\pi_\Upsilon:\Upsilon\To\gr{k}{\tang{M}}$ be the fiber bundle of
the vertical subspaces for the Grassmannian fiber bundle
$\widetilde{\tau}_k:\gr{k}{\tang{M}}\To M$. The fiber at a point
$W\in\gr{k}{\tang{M}}$ is the tangent space of the fiber
$\gr{k}{\tang[x]{M}}$, where $W$ is a subspace of the space
$\tang[x]{M}$ (as it was mentioned early, this space is isomorphic
to the space $Hom\big(W,\,\tang[x]{M}/W\big)$ ).

We have the following exact sequence of the fiber bundles over the
manifold $\gr{k}{\tang{M}}$
\begin{displaymath}
0\;\To\;\Theta\;\stackrel{\ii}{\immerse}\;K\;\stackrel{(\widetilde{\tau}_k)'}{\To}\;\Upsilon\;\To\;0
\end{displaymath}
where $0$ denotes here the trivial fiber bundle with the fibers
equal to $\{0\}$.

In other words, for any $x\in M$ and $W\in\gr{k}{\tang[x]{M}}$, we
have the following exact sequence (see~\ref{exact_sequence1})
\begin{equation}\label{exact_sequence3}
0\;\To\;Hom\big(W,\,\tang[x]{M}/W\big)\;\stackrel{\ii}{\immerse}\;K_W\;\stackrel{(\widetilde{\tau}_k)'(W)}{\To}\;W\;\To\;0
\end{equation}
For any fixed $x\in M$ and $W\in\gr{k}{\tang[x]{M}}$ consider a
submanifold $N\subset{M}$, such that $x\in{N}$ and
$\tang[x]{N}=W$. The submanifold $N$ defines a section
(\textbf{Gauss mapping})
$$g_N:N\To\gr{k}{\tang{M}}|_N$$
where by $\gr{k}{\tang{M}}|_N$ we denote the restriction of the
Grassmann fiber bundle to the submanifold $N$:
$$g_N(x)=\tang[x]{N}\,\subset\,\tang[x]{M}$$
The Gauss mapping $g_N$ induces a mapping $s_N:W\To{K_W}$ which is
a splitting of the exact sequence \ref{exact_sequence3} and is
defined as
\begin{displaymath}
s_N(\xi)=g_N'(x)(\xi)
\end{displaymath}
for any $\xi\in{W}=\tang[x]{N}$. On the other side, this splitting
defines a bilinear form $\scalar{\cdot}{\cdot}_N$ on the vector
space $K_W$ (see the formula \ref{symplectic_form1}), with values
in the quotient space $\tang[x]{M}/W$.
\begin{lem}\label{lemma_submanifold_independence}
The bilinear form $\scalar{\cdot}{\cdot}_N$ is independent of the
choice of the submanifold $N$; i.e., if $N'$ is another
submanifold of the manifold $M$, such that $x\in N'$ and
$\tang[x]{N'}=W$, then the bilinear forms
$\scalar{\cdot}{\cdot}_{N'}$ and $\scalar{\cdot}{\cdot}_N$ are
equal.
\end{lem}
\textbf{Proof.} As it follows from the Lemma
\ref{lemma_splitting_independence}, it is sufficient to prove that
the space $\mathbf{Image}\big(g_{N_1}'(x)\big)$ is isotropic
subspace of the space $K_W$, for the bilinear form
$\scalar{\cdot}{\cdot}_N$.

Using a local linearization of the situation, it is sufficient to
consider the case when $N=F$ and $M=F\times E$, where $F$ and $E$
are finite-dimensional real vector spaces, and
\begin{displaymath}
N_1=\big\{\big(t,f(t)\big)\;|\;\;t\in F,\;x=(0,0)\textrm{ and }f(0)=f'(0)=0\big\}%
\end{displaymath}
In this case we have that the total space of the Grassmannian
fiber bundle is
$$
\gr{k}{\tang{M}}=F\times E\times\gr{k}{F\times E}
$$
the subspace $K_{F\times\{0\}}$ of the vector space
$\tang[(0,0,F)]{F\times E\times\gr{k}{F\times E}}$, corresponding
to the Cartan distribution, is isomorphic to the vector space
$F\times Hom(F\,,\,E)$, and the bilinear form on
$K_{F\times\{0\}}$, corresponding to the submanifold
$F\times\{0\}$ is
\begin{displaymath}
\scalar{(\xi_1,\alpha_1)}{(\xi_2,\alpha_2)}_F=\alpha_1(\xi_2)-\alpha_2(\xi_1)
\end{displaymath}
The Gauss mapping corresponding to the submanifold $N_1$ is
\begin{displaymath}
g_{N_1}(x)=\big\{\big(\xi,\,f'(x)(\xi)\big)\;|\;\;\xi\in F\big\}%
\end{displaymath}
The corresponding splitting at the point $(0,0)\in N_1$
\begin{displaymath}
s_{N_1}\,:\,F\;\To\;F\times Hom(F\,,\,E)
\end{displaymath}
is defined as
\begin{displaymath}
s_{N_1}(\xi)=\big(\xi,\;f''(0)(\xi,\cdot)\big)
\end{displaymath}
for any $\xi\in F$. The image of this mapping is isotropic because
the bilinear mapping $f''(0):F\times F \To E$, corresponding to
the second derivation is always symmetric.
$\Box$

Let the fiber bundle
$$
\widetilde{\tau}_k^{\;*}\big(\tang{M}\big)\To\gr{k}{\tang{M}}
$$
be the pull-back of the tangent bundle of the manifold $M$ to the
total space of the Grassmannization of the tangent bundle.
Consider the fiber bundle
$$
\widetilde{\pi}_\Theta:\widetilde{\Theta}\To\gr{k}{\tang{M}}
$$
which is the quotient of the fiber bundle
$\widetilde{\tau}_k^{\;*}(\tang{M})\To\gr{k}{\tang{M}}$ by the
canonical fiber bundle $\pi_\Theta:\Theta\To\gr{k}{\tang{M}}$ over
the total space of the Grassmanization of the tangent bundle;
i.e., the fiber of the bundle $\widetilde{\pi}_\Theta$ at a point
$W\in\gr{k}{\tang[x]{M}}$ is the quotient space $\tang[x]{M}/W$.
Keeping in mind the Lemma \ref{lemma_submanifold_independence}, we
can state that on the fiber bundle which corresponds to the Cartan
distribution, there is a canonical bilinear form with values in
the fiber bundle
$\tilde{\pi}_\Theta:\tilde{\Theta}\To\gr{k}{\tang{M}}$
\begin{displaymath}
\scalar{\cdot}{\cdot}\;:\;K\oplus{K}\;\To\;\tilde{\Theta}
\end{displaymath}
Let $D:M\To\gr{k}{\tang{M}}$ be a smooth distribution on the
manifold $M$. It is clear that for each point $x\in{M}$, the
mapping
\begin{displaymath}
    D'(x)\;:\;\tang[x]{M}\;\To\;\tang[D(x)]{\gr{k}{\tang{M}}}
\end{displaymath}
carries the subspace $D(x)\subset\tang[x]{M}$ into the space
$K_{D(x)}$ which is the subspace correspondent to the Cartan
distribution, at the point $D(x)\in\gr{k}{\tang{M}}$.
\begin{thm}
The distribution $D:M\To\gr{k}{\tang{M}}$ is integrable if and
only if, for each point $x\in M$, the subspace $D'(x)(D(x))\subset
K_{D(x)}$ is isotropic for the bilinear form
$\scalar{\cdot}{\cdot}:K_{D(x)}\times K_{D(x)}\To\tang[x]{M}/D(x)$
\end{thm}
\textbf{Proof.} For a given distribution $D:M\To\gr{k}{\tang{M}}$ consider
the following bilinear mapping of the fiber bundles
$$
\sigma:D\oplus{D\To\tang{M}/D}
$$
where, for each $x\in M$, the mapping
\begin{displaymath}
\sigma_x:D(x)\times{D(x)\To\tang[x]{M}/D(x)}
\end{displaymath}
is defined as $\sigma_x(u,v)=q([\widetilde{u},\widetilde{v}])$,
where $\widetilde{u}$ and $\widetilde{v}$ are vector fields on the
manifold $M$, such that
$\{\tilde{u},\,\tilde{v}\}\subset{}D,\;\widetilde{u}(x)=u,\;\widetilde{v}(x)=v$
and the mapping
\begin{displaymath}
q\;:\;\tang[x]{M}\;\To\;\tang[x]{M}/D(x)
\end{displaymath}
is the natural quotient mapping.\\
The bilinear form $\sigma_x$ is defined correctly, i.e., the value
$q\big([\tilde{u},\tilde{v}]\big)$ is independent of the choice of
the extensions $\widetilde{u}$ and $\widetilde{v}$. To check this,
consider a vector field $\tau$, on the manifold $M$, such that
$\tau\in D$ and $\tau(x)=0$. Let $\{D_1,\ldots,D_k\}$ be a local
basis of the distribution $D$, and
$\tau=\sum\limits_i\varphi_i{D_i}$, where
$\varphi_i,\;i=1,\ldots,k$ are \smooth- class functions.\\
For any vector field $\xi\in{D}$, we have the following
$$
\begin{array}{c}
[\tau\,,\,\xi]_x=[\sum\limits_i\phi_iD_i\,,\,\xi]_x=\\
\\
=\sum\limits_i\,\underbrace{\phi_i(x)}_0\;[D_i\,,\,\xi]+\sum\limits_i\xi(\phi_i)(x)D_i(x)\,\in\,{D(x)}\then\\
\\
\then q([\tau\,,\,\xi]_x)=0
\end{array}
$$
Using the linearization, introduced in the proof of
Lemma~\ref{lemma_submanifold_independence}, it is easy to see that
the pull-back of the bilinear form $\scalar{\cdot}{\cdot}$ by the
mapping
\begin{displaymath}
D\;:\;M\;\To\;\gr{k}{\tang{M}}
\end{displaymath}
on the subspaces $D(x)\subset\tang[x]{M}$ coincides with the form
$\sigma_x$, i.e., for any $u$ and $v\in{D(x)}$ we have that
$$\sigma_x(u,v)=\scalar{D'(x)(u)}{D'(x)(v)}$$
After this, the statement of the theorem is equivalent to the
Frobenius classical theorem about the integrability of
distributions.
$\Box$

Let $V^k(M),\;k=1,\ldots,\infty$, be the space of antisymmetric,
covariant tensor fields on the manifold $M$, and
$\Omega^k(M),\;k=1,\ldots,\infty$, be the space of differential
$k$-forms on the manifold $M$. Also, we put that
$V^0(M)=\Omega^0(M)=C^\infty(M)$.

If $D$ is a submodule of the $C^\infty(M)$-module $V^1(M)$, then
for any point $x\in M$, we have a subspace of the tangent space of
the manifold $M$ at the point $x$, generated by the set of vectors
$\{\,\xi(x)\;|\;\;\forall\,\xi\in{D}\,\} $ denoted by $D(x)$.
Also, any vector field $\xi\in V^1(M)$, such that
$\xi(x)\in{}D(x)$ for all $x\in M$, is an element of the submodule
$D$. In the case when the dimensions of the subspaces
$D(x)\subset\tang[x]{M},\;x\in M$ are equal to each other, we have
the structure referred as \emph{distribution}, but in some cases
the subspaces $D(x)\subset\tang[x]{M},\;x\in M$ have different
dimensions. In this case, the mapping \mbox{$x\mapsto D(x)$}
is referred as a \emph{singular distribution}.

There is an analogue of the Frobenius theorem for singular
distributions (see \cite{hermann}) which states that the
distribution $D$ (singular or regular) is integrable if and only
if $D$ is involutive and for any vector field $\xi\in D$, the
dimensions of the subspaces $D(x)\subset\tang[x]{M}$ are constant
along the integral paths of the vector field $\xi$.
\newpage
\section{Derivation Based Noncommutative Differential Calculus}
In noncommutative geometry, the commutative algebra of smooth
functions on a smooth manifold is replaced by an abstract algebra,
which, in general, can be noncommutative (see, for example
\cite{connes1}, \cite{dubois1}). The definitions of the classical
geometric objects are translated on the language of the
commutative algebra of the smooth functions and than they are
generalized to the abstract algebra. In this section, we review
the definitions and some facts about diff-geometrical objects on
the language of the noncommutative differential geometry.

\subsection{Noncommutative Differential Forms.}\label{NoncommutativeDiffForms.Section}
Let $A$ be an associative algebra over the field of real or
complex numbers. The space of derivations of the algebra $A$ is
the set of such linear mappings $$X:A\To A$$ that for each $a,b\in
A$:
\begin{displaymath}
X(ab)=X(a)b+aX(b)
\end{displaymath}
It is clear that the space $Der(A)$ is a Lie algebra and if the
algebra $A$ is commutative, then $Der(A)$ is an $A$-module.
Generally, the space $Der(A)$ is a $Z(A)$-module, where $Z(A)$
denotes the center of the algebra $A$.

There are two noncommutative generalizations of the graded
differential algebra of differential forms (see \cite{dubois1},
\cite{dubois2}). The first one is $C_{Z(A)}(Der(A),A)$, which is
the graded algebra of antisymmetric $Z(A)$-multilinear mappings
from $Der(A)$ to $A$. We put that
\begin{displaymath}
    C_{Z(A)}^0(Der(A),A)=A.
\end{displaymath}
The differential operator
\begin{displaymath}
    \ud\;:\;C_{Z(A)}^n(Der(A),A)\;\To\;C_{Z(A)}^{n+1}(Der(A),A)
\end{displaymath}
is defined by the well-known Koszul formula: for any $\omega\in
C_{Z(A)}^n(Der(A),A)$ and $X_1,\ldots,X_{n+1}\in Der(A)$ let
$$
\begin{array}{l}
(\ud\omega)(X_1,\ldots,X_{n+1})=\\
\\
=\sum^{n+1}_{i=1}(-1)^{i+1}X_i\omega(X_1,\ldots,\hat{X_i},\ldots,X_{n+1})+\\
\\
+\sum_{1\leq{i<j}\leq{n+1}}(-1)^{i+j}\omega([X_i,X_j],\ldots,\hat{X_i},\ldots,\hat{X_j},\ldots,X_{n+1})
\end{array}
$$
We denote the space $C_{Z(A)}(Der(A),A)$ by $\Omega_Z(A)$.

The second generalization of the differential forms over the
algebra $A$ is the smallest differential graded subalgebra of the
algebra $\Omega_Z(A)$ containing the algebra $A$. We denote this
algebra simply by $\Omega(A)$. Each element $\omega\in\Omega(A)$
can be expressed as a finite sum of the elements of the type
$a_0\ud{}a_1\cdots{\ud{}a_n}$, where $\ud{}a\in\Omega^1(A)$ is the
one form defined as:
\begin{displaymath}
    (\ud{}a)(X)=X(a),~\textrm{~for~every~}~X\in{Der(A)}.
\end{displaymath}
The multiplication operation in the space $\Omega(A)$ is same as
in the space $\Omega_Z(A)$.

There is a generalization of the classical operator of the inner
derivation $i_X : \Omega^n_Z(A)\To\Omega^{n+1}_Z(A)$ for any
$X\in{Der(A)}$, defined as
$$
\begin{array}{l}
(i_X\omega)(X_1,\ldots,X_{n-1})=\omega(X,X_1,\ldots,X_{n-1})~\textrm{~for~}~\omega\in\Omega^n_Z(A)\\
\textrm{ and }\\
i_X(\alpha)=0 \textrm{ for any }\alpha\in\Omega^0_Z(A)
\end{array}
$$
The subalgebra $\Omega(A)$ is invariant under the action of the
operator $i_X$.

We shall also use, the noncommutative generalization of the
classical Lie derivation operator:
$L_X:\Omega^n(A)\To\Omega^n(A)$, defined as
\begin{displaymath}
L_X=i_X\circ\ud+\ud\circ i_X%
\end{displaymath}
\subsection{Noncommutative Submanifold.}
Let $N$ be a closed submanifold of a smooth compact manifold $M$.
We have the following exact sequence of the commutative algebras
\begin{displaymath}
0\;\To\;I(N)\;\immerse\;C^\infty(M)\;\stackrel{r}{\To}\;C^\infty(N)\;\To\;0
\end{displaymath}
where $r:C^\infty(M)\To C^\infty(N)$ is the restriction mapping
and $I(N)$ is the ideal in the algebra $C^\infty(M)$ consisting of
functions vanishing on the submanifold $N$.

Let $\vecfn{M}$ be the subspace of $\vecf{M}$ consisting of such
vector fields $X$ on the manifold $M$, that
$X(I(N))\subset{I(N)}$. It is clear that if $X\in\vecfn{M}$, then
the restriction of $X$ to the submanifold $N$ is tangent to $N$,
and vice versa: any vector field $\xi\in\vecf{N}$ can be extended
to a vector field $X\in\vecf{M}$, such that $X|_N=\xi$. Therefore,
the restriction mapping
\begin{displaymath}
r\;:\;\vecfn{M}\;\To\;\vecf{N}
\end{displaymath}
is a surjective mapping. The kernel of this mapping is the set of
vector fields on the manifold $M$ vanishing on the submanifold
$N$. In other words
\begin{displaymath}
(X\in\textbf{kernel}(r))\;\same\;(X(C^\infty(M))\subset{I(N)}).
\end{displaymath}
Denote the space $\textbf{kernel}(r)$ by $\vecfn{M}_0$. Hence, we
have the following exact sequence of a Lie algebra homomorphisms
\begin{displaymath}
0\;\To\;\vecfn{M}_0\;\immerse\;\vecfn{M}\;\stackrel{r}{\To}\;\vecf{N}\;\To\;0
\end{displaymath}
To translate these structures on the language of the
noncommutative geometry, consider an associative real or complex
algebra $A$. Let $I$ be an ideal in the algebra $A$. Denote by
$S_I$ the quotient algebra $A/I$ and $q: A \To S_I$ be the natural
quotient mapping.\\
Consider the following Lie subalgebras in $Der(A)$:
$$
\begin{array}{l}
Der_I(A)\;=\;\{\,X\in{Der(A)}\;|\;\;X(I)\subset{I}\,\}\\
\\
Der_I(A)_0\;=\;\{X\,\in{Der(A)}\;|\;\;X(A)\subset{I}\,\}
\end{array}
$$
It is clear that the Lie algebra $Der_I(A)_0$ is an ideal in the
Lie algebra $Der_I(A)$. There is a mapping
$r_I:Der_I(A)\To{Der(S_I)}$, defined as
\begin{equation}\label{VectorFieldRestrictionToSubmanifoldMapping.Formula}
r_I(X)(q(a))\;=\;q(X(a))
\end{equation}
for each $a\in{A}$ and $X\in{Der_I(A)}$. This mapping is the
noncommutative analogue of the restriction mapping
$\vecfn{M}\To\vecf{N}$, which assigns to a vector field on the
manifold $M$, tangent to the submanifold $N$, its restriction to
$N$. The kernel of this mapping is exactly the Lie algebra
$Der_I(A)_0$.

\begin{defn}[see~\cite{masson}]
\label{SubmanifoldAlgebra.Definition} The quotient algebra
$S_I=A/I$ is called a \emph{submanifold algebra} of the algebra
$A$ if the mapping
\begin{displaymath}
    r_I:Der_I(A)\To{Der(S_I)}
\end{displaymath}
is surjective. The ideal $I$ in the algebra $A$ is called the
\emph{constraint ideal} for the submanifold.
\end{defn}
Hence, if the quotient algebra $S_I$ is a submanifold algebra, we
have the following exact sequence of Lie algebra homomorphisms
\begin{displaymath}
0\;\To\;Der_I(A)_0\;\immerse\;Der_I(A)\;\To\;Der(A/I)\;\To\;0
\end{displaymath}
\subsection{Noncommutative Quotient Manifold.}
As before, let $A$ be a real or complex associative algebra and
$B$ be its subalgebra. Consider the following Lie subalgebras of
the Lie algebra $Der(A)$ (see~\cite{masson}):
$$
\begin{array}{l}
Q_B\;=\;\{\,X\in Der(A)\;|\;\;X(B)\subset B\,\}\\
\\
V_B\;=\;\{\,X\in Der(A)\;|\;\;X(B)=0\,\}.
\end{array}
$$
The subalgebra $V_B$ is an ideal in the Lie algebra $Q_B$, i.e.,
$[V_B,Q_B]\subset{V_B}$. We have a natural restriction mapping
\begin{displaymath}
    r_B:Q_B\To{Der(B)}
\end{displaymath}
which is a Lie algebra homomorphism, and the kernel of this
mapping is exactly the Lie algebra $V_B$.
\begin{defn}[see~\cite{masson}]
\label{QuotientManifoldAlgebra.Definition} The subalgebra $B$ of
the algebra $A$ is called a \emph{quotient manifold algebra} of
$A$, if the following conditions are true:
\begin{list}{}{\itemsep=10pt\topsep=15pt}
    \item[\textbf{(q1)}] $Z(B)=B\cap{Z(A)}$
    \item[\textbf{(q2)}] $Der(B)\cong{Q_B/V_B}$
    \item[\textbf{(q3)}] $B=\{a\in A\;|\;X(a)=0,\;\forall\,X\in{V_B}\}$
\end{list}
\end{defn}
Notice, that the condition \textbf{(q1)} is always true if the
algebra $A$ is commutative, and the condition \textbf{(q2)} is
equivalent to the restriction mapping $r_B:Q_B\To Der(B)$ be
surjective. In the latter case we have the following short exact
sequence
\begin{equation}\label{quotient_manifold.exact_seq}
0\;\To\;V_B\;\immerse\;Q_B\;\To\;Der(B)\;\To\;0
\end{equation}
\subsection{Noncommutative Connection and Curvature.}
A bimodule $\Gamma$ over the associative real or complex algebra
$A$, is called a \emph{central bimodule} over the algebra $A$, if
$\Gamma$ is also a \textbf{module} over the center of the algebra
$A$; i.e., for each $s\in\Gamma$ and each $a\in Z(A)$, where
$Z(A)$ is the center of $A$, we have that $as=sa$.

Let a $Z(A)$-module $L$ be a Lie algebra, and we have a Lie
algebra representation of $L$ in the Lie algebra of derivations
$Der(A)$, which is also a $Z(A)$-module homomorphism. Let for any
$z\in{Z(A)}$ and $X,Y\in{L}$, we have that
\begin{equation}\label{LieModule.Formula}
[X\,,\,z\cdot Y]\;=\;X(z)\cdot Y+z\cdot [X\,,\,Y]
\end{equation}
\begin{defn}\label{ConnectionForPair.Definition}
A \emph{connection} for a pair $(L,\;\Gamma)$, where $L$ and
$\Gamma$ are the same as above, is a mapping
$$X\;\mapsto\;\nabla_X$$
where $X\in L$ and $\nabla_X$ is a linear operator
$$\nabla_X\;:\;\Gamma\;\To\;\Gamma$$
satisfying the following conditions:
\begin{list}{}{}
    \item[\textbf{(c1)}] for any $z\in{Z(A)},\;a,b\in{A}$ and $s\in\Gamma$:
        \begin{displaymath}
            \nabla_{(z\cdot X)}(s)=z\cdot\nabla_X(s)
        \end{displaymath}
    \item[\textbf{(c2)}] $\nabla_X(a\cdot s\cdot b)=X(a)\cdot s\cdot b+a\cdot(\nabla_X(s))\cdot b+a\cdot s\cdot X(b)$%
\end{list}
\end{defn}
The mapping $\nabla:L\To{}Hom\big(\Gamma, \Gamma\big)$, is not
necessarily a Lie algebra homomorphism. For any $X,Y\in L$, the
mapping
$$R(X,Y)\;:\;\Gamma\;\To\;\Gamma$$
which measures its deviation from being a homomorphism of Lie
algebras is called the \emph{curvature} of the connection $\nabla$
(see~\cite{dubois3}). More explicitly, for any $X,Y\in{L}$ we have
that
\begin{displaymath}
R(X\,,\,Y)\;=\;[\,\nabla_X\,,\,\nabla_Y\,]-\nabla_{[X\,,\,Y]}
\end{displaymath}
The mapping $R(X,Y):\Gamma\To\Gamma$ is an $A$-bimodule
endomorphism, for any $X,Y\in{L}$.
\subsection{Connection Compatible with Group Action.}
For any subalgebra $B\subset A$, the algebra $A$ is a central
bimodule over the algebra $B$. Let $B$ is a quotient manifold
algebra in the algebra $A$ and $$s:Der(B)\To Q_B$$ be a splitting of
the exact sequence~\ref{quotient_manifold.exact_seq}, which is a
homomorphism of the $Z(B)$-modules, but not necessarily a Lie
algebra homomorphism. Such splitting defines a connection for the
pair $(Der(B)\,,\,A)$ (see~\cite{masson}): for any $X\in Der(B)$,
let the mapping $\nabla_X:A\To A$ be $\nabla_X(a)=s(X)(a)$, for
any $a\in{A}$.

The curvature of this connection is exactly the deviation os the
mapping $s$ from being a Lie algebra homomorphism:
\begin{displaymath}
R_{\nabla}(X\,,\,Y)\;=\;[\,s(X)\,,\,s(Y)\,]-s\big([\,X\,,\,Y\,]\big).
\end{displaymath}
Let $M$ be a smooth manifold and $G$ be a Lie group which acts on
the manifold $M$. Let $A$ be the algebra of smooth functions on
the manifold $M$ and $B$ be the subalgebra of the algebra $A$,
consisting of the functions invariant under the action of the
group $G$. If this action is such that the quotient space $M/G$ is
a smooth manifold, then the subalgebra $B$ is a quotient manifold
algebra of $A$; i.e., for any vector field $X$ on the manifold
$M/G$, exists a vector field $Y$ on the manifold $M$, such that
$q'(Y)=X$, where $q:M\To{M/G}$ is the quotient mapping. As the
mapping $q$ is surjective, the mapping
\begin{displaymath}
q'(x)\;:\;\tang[x]{M}\;\To\;\tang[q(x)]{M/G}
\end{displaymath}
is also surjective for any point $x\in M$. If the quotient space
$M/G$ is a manifold, the vector field $Y$ is a section of the
subbundle of the tangent bundle $\tang{M}$, the fiber of which at
a point $x\in{M}$ is the space $q'(x)^{-1}(X(q(x)))$.

Let $\lie{g}$ be the Lie algebra of the Lie group $G$, and the
mapping
\begin{displaymath}
o\;:\;\lie{g}\;\To\;Der(A)
\end{displaymath}
be the natural Lie algebra homomorphism induced by the action of
the group $G$ on the manifold $M$; i.e., for any $x\in M$ and
$u\in\lie g$, let
\begin{displaymath}
o(u)\;=\;\phi'(1,x)(u,0),
\end{displaymath}
where $\phi:G\times M\To M$ is the mapping defining the group
action. The Lie algebra $V_B$ is the submodule in the $A$-module
$Der(A)$ generated by the subspace $\textbf{Image}(o)$, and the
subalgebra $Q_B$ is the maximal subspace of $Der(A)$ such that
$\big[\textbf{Image}(o),Q_B\big]\subset V_B$ (or, equivalently:
$[V_B\,,\,Q_B]\subset{}V_B$).
\begin{defn}\label{action_compatible_connection.defn}
\emph{A connection compatible with the action of the Lie group}
$G$, is a splitting $s:Der(B)\To Q_B$ of $Z(B)$-modules, such that
for any $X\in{Der(B)}$ and $u\in\lie{g}$, we have that
$[\,o(u)\,,\,s(x)\,]\,=\,0$.
\end{defn}
Now, let us translate this construction on the language of the
noncommutative geometry.

The group action on a manifold can be generalized as a Lie
subalgebra $\lie{g}\subset Der(A)$. The submodule of the vertical
vector fields corresponding to this action can be described as the
$A$-submodule $V^{\lie{g}}\subset Der(A)$ generated by $\lie{g}$.
Let $B^{\lie{g}}$ be the subalgebra of the algebra $A$, such that,
for any $b\in B^{\lie{g}}$ and $v\in V^{\lie{g}}\,:\,v(b)=0$. Let
$Q^{\lie{g}}$ be the maximal subspace of $Der(A)$ such that
$[\,V^{\lie{g}}\,,\,Q^{\lie{g}}\,]\subset V^{\lie{g}}$. It is
clear that $Q^{\lie{g}}$ is a Lie subalgebra of the Lie algebra
$Der(A)$ and is a module over $Z(B^{\lie{g}})$. Recall that
$V_{B^{\lie{g}}}$ is such subspace of $Der(A)$ that
$V_{B^{\lie{g}}}(B^{\lie{g}})=\{0\}$, and $Q_{B^{\lie{g}}}$ is
such that $Q_{B^{\lie{g}}}(B^{\lie{g}})\subset B$.
\begin{lem}\label{qu_in_qub.lemma}
The space $Q^{\lie{g}}$ is a subspace of the space
$Q_{B^{\lie{g}}}$.
\end{lem}
	\textbf{Proof.} For any $\xi\in{Q^{\lie{g}}},\;b\in{B^{\lie{g}}}$ and
$v\in{V^{\lie{g}}}$ we have the following:
\begin{displaymath}
v(\xi(b))\;=\;\underbrace{[v,\xi]}_{\in{V}}\,(b)\;-\;\xi\,(\;\underbrace{v(b)}_0\;)\;=\;0,
\end{displaymath}
therefore: $\xi(b)\in B$.
$\Box$

It is clear that $V^{\lie{g}}\subset{V_{B^{\lie{g}}}}$ but in
general, these two Lie algebras are not identical.
\begin{prop}\label{vg_eq_vbg_follows_qg_eq_qbg.prop}
If $V^{\lie{g}}=V_{B^{\lie{g}}}$ then
$Q^{\lie{g}}=Q_{B^{\lie{g}}}$.
\end{prop}
\textbf{Proof.} We have already proved that $Q^{\lie{g}}$ is always a
subspace of $Q_{B^{\lie{g}}}$. Now the task is to prove that if
$V^{\lie{g}}=V_{B^{\lie{g}}}$ then $Q_{B^{\lie{g}}}$ is a subspace
of the space $Q^{\lie{g}}$. It is equivalent to
$[V^{\lie{g}},Q_{B^{\lie{g}}}]\subset V^{\lie{g}}$. But if
$V^{\lie{g}}=V_{B^{\lie{g}}}$ this statement is the same as
$V^{\lie{g}}$ be a Lie ideal in the Lie algebra $Q_{B^{\lie{g}}}$,
which follows from the fact that $V^{\lie{g}}$ is a kernel of the
Lie algebra homomorphism
$$r_{B^{\lie{g}}}:Q_{B^{\lie{g}}}\To{Der(B_{\lie{g}})}$$
(see the short exact sequence~\ref{quotient_manifold.exact_seq}).
$\Box$

\begin{defn}\label{connection_compat_with_action_noncomm.definition}
In the case when $V^{\lie{g}}=V_{B^{\lie{g}}}$, and the subalgebra
$B^{\lie{g}}$ in the algebra $A$ is a quotient manifold
subalgebra, a connection compatible with the Lie subalgebra
$\lie{g}\subset Der(A)$ is a splitting
$$s:Der(B^{\lie{g}})\To{}Q^{\lie{g}}$$
such that $[\lie{g},s(x)]=0$ for any $x\in Der(B^{\lie{g}})$.
\end{defn}
\subsection{Noncommutative Distribution. Integral Manifold and Bott
Connection.} Further, any submodule of the $Z(A)$-module $Der(A)$,
where $A$ is an associative complex or real algebra, we call a
\emph{distribution}. If a distribution $D$ is a Lie subalgebra of
the Lie algebra $Der(A)$, it will be said to be an
\emph{involutive} distribution.
\begin{defn}\label{NoncommutativeIntegralSubmanifold.Definition}
Let $I$ be an ideal in the algebra $A$, such that the quotient
algebra $S_I=A/I$ is a submanifold algebra (see
Definition~\ref{SubmanifoldAlgebra.Definition}). The quotient
algebra $S_I$ is said to be a \emph{integral submanifold} algebra
for the distribution $D$, if $D$ is is a subalgebra of the Lie
algebra $Der_I(A)$ and $r_I(D)=Der(S_I)$, where
$r_I:Der_I(A)\To{}Der(S_I)$ is defined by the
formula~\ref{VectorFieldRestrictionToSubmanifoldMapping.Formula}.
\end{defn}
In other words, if the ''submanifold`` $A/I$ is integral for the
distribution $D$, then for each $X\in D$, we have that
$X(I)\subset I$; and any ''vector field`` $Y\in Der(A/I)$ can be
extended to the ''vector field`` $X\in D$.

For a given distribution $D$, let us denote by $A_D$ the
subalgebra of the algebra $A$ defined as
\begin{displaymath}
A_D\;=\;\{\;a\;\in\;{A}\;\;|\;\;X(a)=0,\;\forall\,X\in D\;\}
\end{displaymath}
In the classical geometric situation, if $D$ is ivolutive and
regular distribution on some smooth manifold $M$, the algebra
$A_D$ coincides with the subalgebra of the smooth functions on
$M$, constants along the leaves of the foliation, defined by the
distribution $D$. In some ''good`` cases, when the quotient space
of the manifold $M$, by the integral submanifolds of the
distribution $D$ is a smooth manifold, the subalgebra $A_D$ is, in
fact, the algebra of smooth functions on this quotient manifold.
In such situations, the algebra $A_I$ is a \emph{quotient manifold
algebra} in the sense of the noncommutative definition (see
Definition~\ref{QuotientManifoldAlgebra.Definition}).

Let $L$ be a Lie algebra and a module over a commutative algebra
$A$, satisfying the condition \ref{LieModule.Formula}. Let $L_0$
be a Lie subalgebra of the Lie algebra $L$, and le $L_0$ is also a
submodule of the $A$-module $L$. In this situation, the quotient
space $\Gamma=L/L_0$ inherits an $A$-module structure from $L$ and
there is a canonical connection for the pair
$(L_0,~\Gamma=L/L_0)$, defined as
\begin{equation}\label{BottConnection0.Formula}
\nabla_X(q(u))\;=\;q([X\,,\,u])
\end{equation}
for any $X\in{L_0}$ and $u\in{L}$, where $q$ is the natural
quotient mapping, assigning to each $u\in L$, its equivalency
class $q(u)\in L/L_0$. This definition is correct, i.e., for any
two elements $u_1,u_2\in L$, if $q(u_1)\;=\;q(u_2)$ then
$\nabla_X(q(u_1))=\nabla_X(q(u_2))$. To check this, recall that by
definition
\begin{displaymath}
\nabla_X(q(u_1))\;-\;\nabla_X(q(u_2))\;=\;q\big([X\,,\,u_1-u_2]\big),
\end{displaymath}
and the latter is equal to $0$, because: $X,\;u_1-u_2\in L_0$ and
$L_0$ is a Lie subalgebra of $L$.

The mapping $X\mapsto\nabla_X$ satisfies the conditions
\textbf{(c1), (c2)} required for a connection (see
Definition~\ref{ConnectionForPair.Definition}):\\
for any $a\in A$, we have
$$
\begin{array}{l}
\\{}
\nabla_{aX}(q(u))\;=\;q([aX,u])\;=\;q(\,\underbrace{u(a)X}_{\in L_0}\;+\;a[X,u])\;=\\
\\
=\;q(a[X,u])\;=\;aq([X,u])\;=\;a\nabla_X(q(u))
\\{}
\end{array}
$$
and
$$
\begin{array}{l}
\\{}
\nabla_X(aq(u))\;=\;\nabla_X(q(au))\;=\;q([X,au])\;=\\
\\
=\;q(X(a)u)\;+\;q(a[X,u])\;=\;X(a)q(u)\;+\;a\nabla_X(q(u))
\\{}
\end{array}
$$
Now, let us give some kind of dual definition of the connection
described above.

Denote by $\lief$ the space of $A$-valued 1-forms on the
$A$-module $L$, i.e., each element $\alpha\in\lief$ is a mapping
$\alpha:L\To A$, such that for any $X\in L$ and $a\in A$ we have
that $\alpha(aX)=a\alpha(X)$.

For any Lie subalgebra $L_0\in L$, which is also a submodule, let
us denote by $\sublief$ the submodule of the $A$-module $\lief$
consisting of the forms vanishing on the submodule $L_0$.

Define a connection $\nabla$, for the pair $(L_0\,,\,\sublief)$ as
follows:\\
for any $X\in{L_0}$ and $\alpha\in\sublief$, let
\begin{equation}\label{BottConnectionOnForms.Formula}
\nabla_X(\alpha)\;=\;L_X(\alpha)
\end{equation}
where $L_X:\lief\To\lief$ is the operator of Lie derivation:
\begin{displaymath}
L_X\;=\;\ud\circ{i_X}\;+\;i_X\circ{\ud}.
\end{displaymath}
By definition of the space $\sublief$, we have that for any
$\alpha\in\sublief$ and $X\in{L_0}$: $i_X(\alpha)=0$. Therefore,
the formula~\ref{BottConnectionOnForms.Formula} can be simplified
as
\begin{equation}\label{BottConnectionOnFormsSimplified.Formula}
\nabla_X(\alpha)\;=\;\big(i_X\circ{\ud}\big)(\alpha)
\end{equation}
\begin{lem}
For any $X\in{L_0}$, the subspace $\sublief$ in the space $\lief$
is invariant under the action of the operator $\nabla_X$.
\end{lem}
\textbf{Proof.} For $\alpha\in\sublief$ and $X,Y\in{L_0}$, we have
$$
\begin{array}{l}
(\nabla_X(\alpha))(Y)\;=\;((i_X\circ\ud)(\alpha))(Y)\;=\;(\ud\alpha)(X,Y)\;=\\
\\
=\;X\alpha(Y)\;-\;Y\alpha(X)\;-\;\alpha([X,Y])\;=0.
\end{array}
$$
$\Box$

\begin{lem}
The connection $\nabla$ for the pair $(L_0,\sublief)$ is flat
\end{lem}
	\textbf{Proof.} For any $X,Y\in{L_0}$, we have
\begin{displaymath}
R(X,Y)\;=\;[\nabla_X,\nabla_Y]\;-\;\nabla_{[X,Y]}\;=\;[L_X,L_Y]\;-\;L_{[X,Y]}\;=\;0
\end{displaymath}
which follows from the well-known fact that the mapping $X\mapsto
L_X$ is a Lie algebra homomorphism.
$\Box$


To apply the construction described above to the classical
geometric case, consider the case when $L$ is the algebra of
smooth vector fields on a smooth manifold $M$ and
$L_0=D\subset\vecf{M}$ be any involutive distribution on the
manifold $M$. If the distribution $D$ is regular, then as it
follows from the Frobenius theorem, it is integrable. The natural
connection for the pair $(D,\Omega^1_D(M))$, where $\Omega^1_D(M)$
denotes the space of 1-forms vanishing on the vector fields from
$D$, is in fact, the well-known \textbf{Bott connection} for the
foliation of
the integral leaves of the distribution $D$. More precisely:\\
let $N\subset M$ be an integral submanifold of the distribution
$D$. Consider a fiber bundle
$$\pi:E_D\To N$$
the fiber of which at a point $x\in{N}$ is the quotient space
$\tang[x]{M}/\tang[x]{N}$ (or, in the dual terminology: the
subspace of the space $\ctang[x]{M}$ consisting of the 1-forms
vanishing on the subspace $\tang[x]{M}$). The connection for the
pair $(D,\Omega^1_D(M))$ corresponds to some ordinary connection
on the fiber bundle $\pi:E_D\To N$, which is known as the Bott
connection on a leave of a foliation. A strict algebraic
definition of this connection in the terms of noncommutative
submanifolds will be given further.

Let $I\subset A$ be an ideal, invariant under the action of the
Lie subalgebra $L_0\subset L$. That is, for any $X\in L_0$ and
$a\in I$, we have $X(a)\in L$. Introduce the following two spaces:
$$
\begin{array}{l}
\\{}
I\cdot L_0\;=\;\{\,k_1u_1+\cdots+k_nu_n\;|\;\;k_1,\ldots,k_n\in I,\;u_1,\ldots,u_n\in L\,\}\\%
\\
\textrm{and}\\
\\
I\cdot\sublief\;=\\
\\
=\;\{\,k_1\alpha_1+\cdots+k_n\alpha_n\;|\;\;k_1,\ldots,k_n\in
I,\;\alpha_1,\ldots,\alpha_n\in\sublief\,\}
\\{}
\end{array}
$$
\begin{lem}\label{Bott1.Lemma}
For any $X\in L_0$, the submodule $I\cdot\sublief$ is invariant
under the action of the covariant derivation $\nabla_X$
\end{lem}
\textbf{Proof.} For any $X\in L_0,\;a\in I$ and $\alpha\in\sublief$ we have
the following
\begin{displaymath}
\begin{array}{l}
\nabla_X(a\alpha)\;=\;i_X(\ud(a\alpha))\;=\\
\\
=\;i_X((\ud{}a)\alpha\;+\;a(\ud\alpha))\;=\;X(a)\alpha\;+\;a\nabla_X(\alpha).
\end{array}
\end{displaymath}
As the ideal $I$ is invariant under the action of the Lie
subalgebra $L_0$, we have that $X(a)\in{I}$; a submodule
$\sublief$ is invariant under the Lie derivation, we have that
$\nabla_X(\alpha)\in\sublief$. Consequently, we obtain that
$\nabla_X(a\alpha)\in I\cdot\sublief$.
$\Box$

It is easy to check that the submodule $I\cdot L_0$ is a Lie ideal
in the Lie algebra $L_0$, that is, for any
$a\in{}I,\;u,v\in{}L_0:\;[au,\,v]\in{}L_0$. As $I\cdot\sublief$ is
invariant under the action of the Lie algebra $L_0$, we can define
the action of $L_0$ on the quotient module
\begin{displaymath}
\sublieff.
\end{displaymath}
\begin{lem}\label{Bott2.Lemma}
The action of the Lie subalgebra $I\cdot L_0\subset L_0$ on the
quotient module $\sublieff$ is trivial.
\end{lem}
\textbf{Proof.} It is equivalent to the statement that any element of the
Lie algebra $I\cdot L_0$ carries the elements of the space
$\sublief$ in the subspace $I\cdot\sublief$.

Consider $a\in I,\;u\in L_0$ and $\alpha\in\sublief$. For these
elements, we have the following
\begin{displaymath}
\nabla_{(au)}(\alpha)\;=\;i_{(au)}(\ud\alpha)\;=\;ai_u(\ud\alpha)\;=\;a\nabla_u(\alpha)\;\in\;I\cdot\sublief.
\end{displaymath}
$\Box$

As the submodule $I\cdot L_0$ is a Lie algebra ideal in $L_0$, we
have that the quotient module $L_0/I\cdot L_0$ is a Lie algebra
and it follows from the Lemma~\ref{Bott1.Lemma} and
\ref{Bott2.Lemma}, that the natural connection on the pair
$(L_0,\sublief)$ induces a connection on the pair
\begin{displaymath}
    \big(\,L_0/I\cdot L_0\,,\,\sublieff\big)%
\end{displaymath}
In the case of the classical geometry, the subalgebra $L_0$
corresponds to an involutive distribution $D$ on a smooth manifold
$M$; $I$ is an ideal of smooth functions on the manifold $M$,
corresponding to some integral submanifold of the distribution
$D$. The quotient Lie algebra $L_0/I\cdot L_0$ can be identified
with the Lie algebra of the vector fields on the submanifold $N$;
and the quotient module $\sublieff$ --- to the module of the
sections of the fiber bundle $\pi : E_D\To N$.

Now, we describe another algebraic model for the Bott connection
on an integral submanifold.

Let $D\subset Der(A)$ be a distribution (i.e., a Lie subalgebra)
and $I$ be an ideal in the algebra $A$, such that the quotient
algebra $A/I$ is an integral submanifold algebra for the
distribution $D$. As it follows from the definition of the
noncommutative integral submanifold (see
Definition~\ref{NoncommutativeIntegralSubmanifold.Definition}), we
have the following short exact sequence
\begin{equation}\label{IntegralSubmanifoldExactSequence.Formula}
0\;\To\;D_I\;\immerse\;D\;\stackrel{r_I}{\To}\;Der(A/I)\;\To\;0
\end{equation}
where $D_I$ is the subalgebra of the Lie algebra $D$, that can be
described as
\begin{displaymath}
D_I\;=\;\{\,X{\in}D\;\;|\;\;X(A)\subset I\,\}
\end{displaymath}
Let $\formsD$ be the submodule of the module $\Omega^1(A)$
consisting of the 1-forms vanishing on the vector fields from the
distribution $D$. As it follows from our previous discussion,
there is a natural connection for the pair $(D,\;\formsD)$. Let us
give an algebraic description of the Bott connection on an
integral submanifold of a distribution.

Let $\formsI$ be the submodule of $\Omega^1(A)$, consisting of
such forms $\alpha$, that for any
$X\in{}Der(A):\;\alpha(X)\in{}I$.
\begin{rem}
In the sense of the classical geometry, the space $\formsI$
corresponds to the submodule of the forms vanishing on the
submanifold constrained by the ideal $I$.
\end{rem}
Denote by $\Gamma(D,I)$ the quotient module
$\formsD/(\formsD\cap\formsI)$.

For each $U\in D/D_I\cong Der(A/I)$ define an operator
\begin{displaymath}
\nabla_U\;:\;\Gamma(D,I)\;\To\;\Gamma(D,I)
\end{displaymath}
as
\begin{displaymath}
\nabla_U(q_2(\alpha))\;=\;q_2(L_X(\alpha))
\end{displaymath}
where $u=q_1(x)$ and the mappings
\begin{displaymath}
\begin{array}{l}
q_1\;:\;D\;\To\;D/D_I\\
\\
q_2\;:\;\formsD\;\To\;\formsD/(\formsD\cap\formsI)%
\end{array}
\end{displaymath}
are the canonical projection mappings.
\begin{prop}
The correspondence $U\mapsto\nabla_U$ is a connection for the pair
$(\,Der(A/I)\,,\,\Gamma(D,I)\,)$.
\end{prop}
\textbf{Proof.} It is sufficient to check that the definition of the
operator $\nabla_U$ is \textbf{correct}. For this, we should check
that for any $V\in D_I$ and
$\alpha{}\in\formsD:\;L_V(\alpha)\in\formsI$. By definition, we
have that for any element $X\in Der(A)$
\begin{displaymath}
\begin{array}{l}
\big(L_V(\alpha)\big)(X)\;=\;\big((i_V{\circ}\ud\big)(\alpha))(X)\;=\\
\\
=\;V\alpha(X)\;-\;X\alpha(V)\;-\;\alpha([V,X])
\end{array}
\end{displaymath}
From the definition of the Lie ideal $D_I$ follows that
$V\alpha(X)\in I$.\\
From the definition of $\formsD$, follows that $\alpha(V)=0$.\\
To show that $\alpha\big([V,X]\big)\in I$, recall that the form
$\alpha$ can be represented as $\alpha=\sum\limits_i{}a_i\ud{}b_i$
where $a_i,b_i\in A$. Therefore
\begin{displaymath}
\begin{array}{c}
\alpha([V,X])\;=\;V(\sum\limits_ia_iX(b_i))\;-\;X\alpha(V)\;=\\
\\
=\;V(\sum\limits_ia_iX(b_i))\;\in\;I%
\end{array}
\end{displaymath}
which follows from the fact that $\alpha(V)=0$ and $V\in D_I$.
$\Box$

\begin{defn}
The connection for the pair $(\,Der(A/I)\,,\,\Gamma(D,I)\,)$ is
called the \textbf{Bott connection for the integral manifold
constrained by the ideal} $I$.
\end{defn}
\newpage
\section{Lie Superalgebra Structures on the Space of Multiderivations}
\subsection{Compositional Product and Supercommutator}
Let $V$ be a vector space over the field of real or complex
numbers. For each integer $k\geq 0$ let us denote by $L^k(V)$ the
space of multilinear antisymmetric mappings from $V^k$ into $V$.
We also set that
$$
L^0(V)=V \textrm{ and }L(V)={\bigoplus\limits_{k=0}^\infty}L^k(V)%
$$
There is a natural associative algebra structure on the space
$L^1(V)$, defined by the composition of the elements as linear
operators, and a natural Lie algebra structure induced by the
above associative algebra structure, where the Lie bracket is the
ordinary commutator of two linear operators:
$$[\alpha,\beta]\;=\;\alpha\circ\beta\;-\;\beta\circ\alpha$$ for
all ${\alpha,\beta\in}L^1(V)$.

The multiplication operation (composition) on the space $L^1(V)$
can be extended to the space $L(V)$ and the resulted operator is
called the \textbf{compositional product} (see~\cite{nijenhuis}):
for $\alpha\in L^m(V)$ and $\beta\in L^n(V)$ the compositional
product $\alpha\circ\beta\in L^{m+n-1}(V)$ is
\begin{equation}\label{CompositionalProduct.Formula}
\begin{array}{l}
    (\alpha\circ\beta)(a_1,\ldots,a_{m+n-1})=\\
    \\
    =\sum\limits_s{\textrm{sgn}(s)\,\alpha(\beta(\;\underbrace{a_{s(1)},\ldots,a_{s(n)}}_{s(1)<\cdots<s(n)}\,),\;\underbrace{a_{s(n+1)},\ldots,a_{s(m+n-1)}}_{s(n+1)<\cdots<s(m+n-1)}\;)}%
\end{array}
\end{equation}
where $a_1,\ldots,a_{m+n-1}$ are elements of the vector space $V$
and $\textrm{sgn}(s)$ is the signature of the permutation $s$.

As a result, the commutator existing on the space $L^1(V)$ can be
extended to the operation on the space $L(V)$, called a
\textbf{supercommutator}:
\begin{equation}\label{Supercommutator0.Formula}
[\,\alpha\,,\,\beta\,]=(-1)^{(m+1)n}\,\alpha\circ\beta\,+\,(-1)^m\,\beta\circ\alpha
\end{equation}
This operation (bracket) satisfies the following conditions, which
makes the space $L(a)$ a Lie superalgebra: for
$\alpha\in{}L^m(V),\;\beta\in{}L^n(V)$ and $\gamma\in L^k(V)$
\begin{list}{}{\topsep=15pt\itemsep=15pt}\label{SuperalgebraConditions}
    \item[\textbf{(s1)}] $[\,\alpha\,,\,\beta\,]=(-1)^{mn}[\,\beta\,,\,\alpha\,]$;
    \item[\textbf{(s2)}] $(-1)^{mk}[\,[\,\alpha\,,\,\beta\,]\,,\,\gamma\,]+(-1)^{mn}[\,[\,\beta\,,\,\gamma\,]\,,\,\alpha\,]+(-1)^{nk}[\,[\,\gamma\,,\,\alpha\,]\,,\,\beta\,]=0$
\end{list}
\bigskip
An element $\mu\in L^2(V)$ satisfying the condition $[\mu,\mu]=0$,
we call an \emph{involutive} element. Such element defines a Lie
algebra structure on the vector space $V$, via the commutator
\[\,[\,a\,,\,b\,]_{\mu}\;=\;\mu(\,a\,,\,b\,)\]
for $a,b\in V$. Notice, that the Jacoby identity for the
commutator $[\,\cdot\,,\,\cdot\,]_{\mu}$ is equivalent to the
condition $[\mu,\mu]=0$.
Any involutive element $\mu\,\in\,L^2(V)$ defines a linear opeator
$$\cd[\mu]\,:\,L(V)\;\To\;L(V)$$
as
$$\cd[\mu](\alpha)\;=\;[\,\mu\,,\,\alpha\,]$$
for any $\alpha\in L(V)$. The degree of this operator is equal to
$+1$, that is
$$\cd[\mu]\big(L^k(V)\big)\;\subset\;L^{k+1}(V)$$
for any integer $k\geq 0$.

From the condition \textbf{(s2)} for the commutator
$[\,\cdot\,,\,\cdot\,]$ and the fact that the element
$\mu\in{}L^2(V)$ is involutive, easily follows that the operator
\cd[\mu] is a coboundary operator: $\cd[\mu]\circ\cd[\mu]=0$.
\subsection{Supercommutator on the Space of Multiderivations}
Now, let $V=A$ be a real or complex commutative algebra. In this
case, the space is endowed with a structure of exterior algebra
under the multiplication operation defined by the classical
formula. For $\alpha\in~L^m(A)$ and $\beta\in~L^n(A)$ we have that
\begin{equation}\label{ExteriorMultiplication.Formula}
\begin{array}{l}
\big(\alpha\wedge\beta\big)(u_1,\ldots,u_{m+n})\;=\\
\\
=\;\frac{1}{m!n!}\sum\limits_s\textrm{sgn}(s)\alpha(u_{s(1)},\ldots,u_{s(m)})
    \beta(u_{s(m+1)},\ldots,u_{s(m+n)})\\
\end{array}
\end{equation}
where $\{u_1,\ldots,u_{m+n}\}\subset A$.
\begin{defn}
an element $\alpha\in L(A)$ is said to be a \emph{multiderivation}
if for any set of elements $\{a,a_1,\ldots,a_k\}\subset A$ we have
that
\begin{displaymath}
\alpha(aa_1,a_2,\ldots,a_k)\;=\;a\alpha(a_1,\ldots,a_k)\;+\;a_1\alpha(a,a_2,\ldots,a_k)
\end{displaymath}
For any integer $k\geq 0$ the subspace of all multiderivations in
the space $L^k(A)$ we denote by $Der^k(A)$. Also, we set that
$Der^0(A)=A$ and $Der(A)=\bigoplus\limits_k^{\infty}Der^k(A)$.
\end{defn}
The subspace $Der(A)$ in the space $L(A)$ is closed as under the
operation of the exterior multiplication defined by the
formula~\ref{ExteriorMultiplication.Formula}, so under the
supercommutator defined by the
formula~\ref{Supercommutator0.Formula}. Moreover, these two
structures are interconnected via the following property
\begin{equation}\label{Superbiderivation.Formula}
[\,\alpha\,,\,\beta\gamma\,]\;=\;[\,\alpha\,,\,\beta\,]\wedge\gamma\;+\;(-1)^{(m+1)n}\,\beta\wedge[\,\alpha\,,\,\gamma\,]
\end{equation}
for any $\alpha\in Der^m(A)\;,\;\beta\in Der^n(A)$ and $\gamma\in
Der(A)$.

In the classical case, when $A=C^{\infty}(M)$, the supercommutator
$[\,\cdot\,,\,\cdot\,]$ on the space $Der(A)$ is called the
\textbf{Schouten bracket}.

For any integer $k\geq 0$ consider $\wedge^kDer^1(A)$ which is the
subspace of the space $Der^k(A)$, consisting of the linear
combinations of the elements of the type
$v_1\wedge\cdots\wedge{}v_k,\;\{v_1,\ldots,v_k\}\subset{}Der^1(A)$.
The space
$$
{\wedge}Der^1(A)=\bigoplus_{k=0}^{\infty}\wedge^kDer^1(A)
$$
is a subalgebra in the exterior algebra $Der(A)$ and also is
closed under the supercommutator. The explicit formula for the
restriction of the supercommutator on the subspace $\wedge
Der^1(A)$ is the following:
\begin{equation}\label{SupercommutatorInCoordinates.Formula}
\begin{array}{l}
[\,u_1\wedge\ldots\wedge{u_m}\,,\,v_1\wedge\ldots\wedge{v_n}\,]\;=\;\\
\\
=\;\sum\limits_{i,j}(-1)^{m+i+j-1}[\,u_i\,,\,v_j\,]\wedge u_1\wedge\cdots\wedge\hat{u_i}\wedge\cdots\wedge u_m\wedge\\
\wedge{v_1\wedge\cdots\wedge\hat{v_j}\wedge\cdots\wedge v_n}
\end{array}
\end{equation}
where $u_1,\ldots,u_m$ and $v_1,\ldots,v_n$ are the elements of
the space $Der^1(A)$.
\subsection{Poisson Structure. Poisson Cohomologies}
As it was mentioned early, an involutive element $P\in L^2(A)$
defines a commutator $[\,\cdot\,,\,\cdot\,]_P:A\times A\To A$. In
the case when $P\in Der^2(A)$, this commutator is a biderivation.
That is: for any $a,b \textrm{ and }c\in A$, we have
$$
[\,a\,,\,b\cdot c\,]\;=\;b\cdot[\,a\,,\,c\,]+[\,a\,,\,b\,]\cdot c
$$
\begin{defn}\label{PoissonAlgebra.Definition}
A \emph{Poisson algebra} is a real or complex vector space $A$,
with two operations
$$
(a,b)\mapsto ab\qquad\textrm{ and }\qquad(a,b)\mapsto\{a,b\}
$$
satisfying the following three conditions:
\begin{list}{}{\leftmargin=15pt\topsep=15pt\itemsep=15pt}
\item[\textbf{(p1)}]the operation (multiplication) $(a,b)\mapsto ab$
    makes $A$ an associative algebra;
\item[\textbf{(p2)}]the operation (bracket) $(a,b)\mapsto\{a,b\}$
    makes $A$ a Lie algebra;
\item[\textbf{(p3)}]these two operations are related via the Leibnitz
    rule: $$\{\,a\,,\,b\cdot c\,\}\;=\;b\cdot\{\,a\,,\,c\,\}+\{\,a\,,\,b\,\}\cdot c$$
\end{list}
\end{defn}
So, it can be stated that, an involutive element $P\in Der^2(A)$
defines a Poisson algebra structure on the commutative algebra
$A$: $$\{\,a\,,\,b\,\}_P\;=\;P(\,a\,,\,b\,)$$

As the subspace $Der(A)\subset L(A)$ is closed under the
supercommutator, it will be invariant under the action of the
coboundary operator \cd[P]. Therefore, we have the subcomplex
$\big(Der(A)\,,\,\cd[P]\big)$ of the complex
$\big(L(A)\,,\,\cd[P]\big)$.

From the condition~\ref{Superbiderivation.Formula}, for the
commutator $[\,\cdot\,,\,\cdot\,]$ on the exterior algebra
$Der(A)$, follows that the operator $\cd[P]:Der(A)\To Der(A)$ is
an antidifferential, that is: for any $u\in Der^m(A)$ and $v\in
Der^n(A)$
$$
\cd[P](u\wedge v)\;=\;\cd[P](u)\wedge v\;+\;(-1)^mu\wedge\cd[P](v)
$$
Therefore, the exterior algebra structure on the space $Der(A)$,
induces an exterior algebra structure on the cohomologies of the
complex $(Der(A)\,,\,\cd[P])$.

The cohomology algebra of the complex $(Der(A)\,,\,\cd[P])$, is
said to be the \emph{cohmologies of the Poisson structure} $(A,P)$
or simply, \emph{the Poisson cohomologies}.
\subsection{External Differential as a Supercommutator}
For a real or complex vector space $V$, let us denote by
$\mend{V}$ the space of linear endomorphisms of the space $V$. If
$V$ is a module over the commutative algebra $A$, we denote by
$\mend[A]{V}$ the space of $A$~-~modular endomorphisms of the
space $V$. It is clear that $\mend[A]{V}\subset\mend{V}$. For any
element $a\in A$, the operator of the multiplication on the
element $a$ ($v\mapsto a\cdot v$), we denote by \op{a}.

An element $\phi\in\mend{V}$ is said to be a \emph{first order
differential operator} on the $A$-module $V$, if for any $a\in A$
we have that $[\phi,\op{a}]\in\mend[A]{V}$. The space of first
order differential operators on the $A$-module $V$ we denote by
$\diff{V}$ (see~\cite{vinogradov}). The space $\diff{V}$ is an
$A$-module.

It is clear that for the commutative algebra $A$, the mapping
$a\mapsto\op{a}$ gives an isomorphism $A\iso\mend[A]{A}$.
\begin{lem}\label{Diff(A)andDer(A)plusA.lemma}
The module $\diff{V}$ is canonically isomorphic to the module
$Der^1(A)\oplus A$.
\end{lem}
\textbf{Proof.} Any element $\phi\in\diff{A}$ can be represented as
\[\phi=(\phi-\op{\phi(1)})+\op{\phi(1)}\]
As it follows from the definition of a first order differential
operator, for any $a\in A$, the operator $[\phi,\op{a}]$ is an
element of the space $\mend[A]{A}$. Therefore, for any $b\in A$,
we have \[[\phi,\op{a}](b)=b[\phi,\op{a}](1)\] expanding the last
equality, we obtain \[\phi(ab)+ab\phi(1)=a\phi(b)+\phi(a)b\]
subtracting from the last equality the term $2\cdot\phi(1)ab$, we
easily obtain the equality
\[(\phi-\op{\phi(1)})(ab)=a\cdot(\phi-\op{\phi(1)})(b)+(\phi-\op{\phi(1)})(a)\cdot b\]
which means that the operator $\phi-\op{\phi(1)}:A\To A$ is a
derivation.
$\Box$

The superalgebra structure defined at the beginning of this
section, on the space $L(V)$, where $V$ is a real or complex
vector space can be considered in the case when $V=\diff{A}$. The
space $\diff{A}$ is equipped with the natural structure of a Lie
algebra defined by the commutator: $[u,v]=u\circ v-v\circ u$, for
$u,v\in\diff{A}$, which means that, there exists an involutive
element $\mu\in L^2(\diff{A})$, such that $[u,v]=\mu(u,v)$. The
element $\mu$, itself, defines the coboundary operator
$$
\cd[\mu]\,=\,[\mu,\,\cdot\,]\,:\,L(\diff{A})\;\To\;L(\diff{A})
$$
So, we can talk about the differential complex
$\big(\,L(\diff{A})\,,\,\cd[\mu]\,\big)$.

For each integer $n\geq 1$, let $\widetilde{L}^n(\diff{A})$ be the
subspace of the space $L^n(\diff{A})$, consisting of the mappings
$$
\omega\,:\,\underbrace{\diff{A}\times\cdots\times\diff{A}}_{n-times}\;\To\;\diff{A}
$$
satisfying the following conditions:
\begin{itemize}
\item
the values of the mapping $\omega$ are in $A$ (recall that the
space $\diff{A}$ is isomorphic to (see
Lemma~\ref{Diff(A)andDer(A)plusA.lemma}) ${Der^1(A)\oplus}A$);
\item
$\omega(u_1,\ldots,u_n)=0$ if at least one of the elements
$\{u_1,\ldots,u_n\}\subset\diff{A}$ is in $A$;
\item
$\omega$ is $A$-multilinear mapping; i.e.
\[\omega(a{\cdot}u_1,\ldots,u_n)=a\cdot\omega(u_1,\ldots,u_n)\] for
any $a\in A$ and any collection
$\{u_1,\ldots,u_n\}\subset\diff{A}$.
\end{itemize}
Also, we set that $\widetilde{L}^0(\diff{A})=A$.

It is easy to see that the space $\widetilde{L}^n(\diff{A})$ is
the same as the space of derivation based differential forms for
the commutative algebra $A$: $C^n_{Z(A)}(Der(A),A)$ (see
Section~\ref{NoncommutativeDiffForms.Section}).
\begin{prop}
The subspace
$\widetilde{L}(\diff{A})=\bigoplus_{n=0}^\infty\widetilde{L}^n(\diff{A})$
in the space $L(\diff{A})$ is invariant under the action of the
operator $\cd{\mu}~=~[\mu~,~\cdot]$ and the restriction of the
operator $\cd{\mu}$ on the subalgebra $\widetilde{L}(\diff{A})$,
coincides with the classical \textbf{Koszul} differential on the
space of differential forms (see
Section~\ref{NoncommutativeDiffForms.Section}).
\end{prop}
\textbf{Proof.} By definition of the compositional product and the
corresponding supercommutator (see
Formula~\ref{Supercommutator0.Formula} and
Formula~\ref{CompositionalProduct.Formula}) we have the following
$$
\begin{array}{l}
[\mu,\omega](u_1,\ldots,u_{n+1})\;=\\
\\
=\;(-1)^n\sum(-1)^{n+1-i}\mu(\omega (u_1,\ldots,\widehat{u}_i,\ldots,u_{n+1}),u_i)\;+\\
\\
\qquad+\;\sum(-1)^{i+j-1}\omega(\mu(u_i,u_j),u_1,\ldots,\widehat{u}_i,\ldots,\widehat{u}_j,\ldots,u_{n+1})\;=\\
\\
=\;\sum(-1)^{i-1}[\omega(u_1,\ldots,\widehat{u}_i,\ldots,u_{n+1}),u_i]\;+\\
\\
\qquad+\;\sum(-1)^{i+j-1}\omega([u_i,u_j],\,u_1,\ldots,\widehat{u}_i,\ldots,\widehat{u}_j,\ldots,u_{n+1})\;=\\
\\
=\;\sum(-1)^iu_i\omega (u_1,\ldots,\widehat{u}_i,\ldots,u_{n+1})\;+\\
\\
\qquad+\;\sum(-1)^{i+j-1}\omega([u_i,\,\,\,u_j],\,u_1,\ldots,\widehat{u}_i,\ldots,\widehat{u}_j,\ldots,u_{n+1})\;=\\
\\
=\;-\sum(-1)^{i-1}u_i\omega (u_1,\ldots,\widehat{u}_i,\ldots,u_{n+1})\;+\\
\\
\qquad+\;\sum(-1)^{i+j}\omega([u_i,\,\,\,u_j],\,u_1,\ldots,\widehat{u}_i,\ldots,\widehat{u}_j,\ldots,u_{n+1})\;=\\
\\
=\;-(d\omega)((u_1,\ldots,u_{n+1}))
\end{array}
$$
$\Box$

To summarize, we can state that generally, the subspace
$\widetilde{L}(\diff{A})$ in the space $L(\diff{A})$ is not closed
under the operation of supercommutator, but it is invariant under
the action of the operator $[\mu,\,\cdot]$, where
$\mu\in{}L^2(\diff{A})$ is the element, corresponding to the Lie
algebra commutator in the space $\diff{A}$. It can be defined an
operation of external multiplication in the space
$\widetilde{L}(\diff{A})\iso C^n_{Z(A)}(Der(A),A)$ by the
formula~\ref{ExteriorMultiplication.Formula}, after which, the
operator $\cd{\mu}$ becomes the antiderivation of degree $+1$ on
the algebra $\widetilde{L}(\diff{A})$.

For simplicity, further we denote the space
$\widetilde{L}(\diff{A})$ by $\subforms{A}$.

Any element $p\in Der^2(A)$ defines the mapping
$$\widetilde{p}:A\To{}Der^1(A)$$
as follows
$$
\widetilde{p}(a)(x)\;=\;p(a,x)
$$
which can be extended to the mapping
$$\widetilde{p}:\subforms{A}\To{}Der(A)$$
by the following formula
\begin{equation}\label{tlldeBivectorFieldMapping.Formula}
\widetilde{p}(\alpha)(a_1,\ldots,a_n)\;=\;(-1)^n\alpha\big(\tilde{p}(a_1),\ldots,\tilde{p}(a_n)\big)
\end{equation}
where $\alpha\in\subforms[n]{A}$ and
$\{a_1,\ldots,a_n\}\subset{}A$.

As it was mentioned, the bracket
$$
\{\cdot\,,\,\cdot\}:A\times A\To A,\quad\{a,b\}=p(a,b)
$$
is a Lie algebra structure on $A$ if and only if the element $p$
is involutive ($[\,p\,,\,p\,]\,=\,0$), because of the following
equality
\begin{displaymath}
\frac{1}{2}\,[\,p\,,\,p\,](\,a\,,\,b\,,\,c)\;=\;\{\,\{\,a\,,\,b\,\}\,,\,c\,\}\;+\;\{\,\{\,b\,,\,c\,\}\,,\,a\,\}\;+\;\{\,\{\,c\,,\,a\,\}\,,\,b\,\}
\end{displaymath}
Furthermore, there is true the following
\begin{lem}
If the element $p\in Der^2(A)$ is involutive, the mapping
$$\widetilde{p}:A\To Der^1(A)$$ is a Lie algebra homomorphism.
\end{lem}
\textbf{Proof.} The proof easily follows from the following sequence of
equalities
$$
\begin{array}{l}
\\{}
\widetilde{p}(\{a,b\})(c)\;=\;\{\,\{\,a\,,\,b\,\}\,,\,c\,\}=\{\,a\,,\,\{\,b\,,\,c\,\}\,\}\;-\;\{\,b\,,\,\{\,a\,,\,c\,\}\,\}=\\
\\
=\;\big(\widetilde{p}(a)\widetilde{p}(b)\;-\;\widetilde{p}(b)\widetilde{p}(c)\big)(c)
\end{array}
$$
$\Box$

The following theorem extends the above homomorphism to higher
order elements
\begin{thm}
The mapping $\widetilde{p}:\subforms{A}\To Der(A)$ is a
homomorphism of the differential complexes
$\big(\subforms{A}\,,\,d\big)$ and
$\big(Der(A)\,,\,\cd[p]=[p\,,\,\cdot]\big)$.
\end{thm}
\textbf{Proof.}
$$
\begin{array}{l}
\widetilde{p}(d\omega)(a_1,\ldots,a_{n+1})=\\
\\
=(-1)^{n+1}(d\omega)(\widetilde{p}(a_1),\ldots,\widetilde{p}(a_{n+1}))=\\
\\
=(-1)^{n+1}(\sum_i(-1)^{i-1}\widetilde{p}(a_i)\omega(\widetilde{p}
(a_1),\ldots,\widehat{\widetilde{p}(a_1)},\ldots,\widetilde{p}(a_{n+1}))+\\
\\
+\sum_{i<j}(-1)^{i+j}\omega([\widetilde{p}(a_i),\widetilde{p}
(a_j)],\ldots,\widehat{\widetilde{p}(a_i)},\ldots,
\widehat{\widetilde{p}(a_j)},\ldots,\widetilde{p}(a_{n+1})))=\\
\\
=(-1)^{n+1}(\sum_i(-1)^{i-1}p(a_i,\omega(\widetilde{p}(a_1),\ldots
,\widehat{\widetilde{p}(a_1)},\ldots,\widetilde{p}(a_{n+1}))+\\
\\
+\sum_{i<j}(-1)^{i+j}\omega ([\widetilde{p}(a_i),\widetilde{p}
(a_j)],\ldots,\widehat{\widetilde{p}(a_i)},\ldots,\widehat{\widetilde{p}
(a_j)},\ldots,\widetilde{p}(a_{n+1})))
\\{}
\end{array}
$$
On the other hand we have:
$$
\begin{array}{l}
\\{}
[p,\widetilde{p}(\omega )](a_1,\ldots,a_{n+1})=\\
\\
=\sum_i(-1)^{i-1}p((\widetilde{p}(\omega))(a_1,\ldots,\widehat{a_i},\ldots,a_{n+1}),a_i)+\\
\\
+\sum_{i<j}(-1)^{i+j-1}\widetilde{p}(\omega)(p(a_i,a_j),\ldots,\widehat{a_i},\ldots,\widehat{a_j},\ldots,\widetilde{p}(a_{n+1}))=\\
\\
=(-1)^{n+1}(\,\sum_i(-1)^{i-1}p(a_i,\omega(\widetilde{p}(a_1),\ldots,\widehat{\widetilde{p}(a_i)},\ldots,\widetilde{p}(a_{n+1}))\,)+\\
\\
+\sum_{i<j}(-1)^{i+j-1}\omega([\widetilde{p}(a_i),\widetilde{p}(a_j)],\ldots,\widehat{\widetilde{p}(a_i)},\ldots,\widehat{\widetilde{p}(a_j)},\ldots,\widetilde{p}(a_{n+1})))
\end{array}
$$
$\Box$

\newpage
\section{Schouten Bracket as the Deviation of the\\
Coboundary Operator from the Leibnitz Rule}
\subsection{An Invariant Formula for Schouten Bracket}
The main result of the previous section is the fact that a
supercommutator and a second order involutive element on an
external algebra, defines some coboundary operator on this
algebra; and even \textbf{the classical Koszul differential on the
external algebra of differential forms can be represented as a
supercommutator with some second order element} of some
superalgebra containing the algebra of differential forms. In this
section, we consider some reversed situation: \textbf{a coboundary
operator on some external algebra induces a supercommutator on
this algebra}.

Let $E=\bigoplus_{n=0}^{\infty}E_n$ be a real or complex
$\mathbf{Z}$-graded external algebra with a multiplication
operator denoted by $\wedge$. That is: for $\alpha\in E_m$ and
$\beta\in E_m$, we have that $\alpha\wedge\beta\in E_{m+n}$ and
$\alpha\wedge\beta=(-1)^{m+n}\beta\alpha$, where
$\{m,n\}\subset\mathbf{N}$.

Let $\cd:E\To E$ be a boundary operator; i.e., for any
$k=1,\ldots,\infty$:
$$\cd(E_k)\subset E_{k-1}\textrm{ and }\cd\circ\cd=0.$$

The operator \cd{} is said to be an \emph{antidifferential} if for
any $u\in E_m$ and $v\in E_n$, satisfies the following condition
$$
\cd(u\wedge v)\;=\;\cd(u)\wedge v\;+\;(-1)^mu\wedge\cd(v)
$$
For any boundary operator \cd{} on the external algebra $E$ can be
defined the bilinear mapping
$[\,\cdot\,,\,\cdot\,]:E\times{}E\To{}E$ as
\begin{equation}\label{CommutatorAsDeviation.Formula}
[\,u\,,\,v\,]\;=\;\cd(u)\wedge v\;+\;(-1)^mu\wedge\cd(v)\;-\;\cd(u\wedge v)%
\end{equation}
If the operator \cd{} is an antidifferential, then the commutator
defined by this formula is trivial one: $[u,v]=0$ for any $u,v\in
E$.

In any case, the question: \textbf{is the commutator
$[\cdot,\cdot]$ a Lie superalgebra structure on the graded algebra
$E$ or not?} is natural. To be so, recall that the following
conditions must be hold: for any $u\in E_m,\;v\in E_n$ and $w\in
E_k$
\begin{list}{}{\topsep=15pt\itemsep=10pt}
\item[\textbf{(sa1)}] $[u,v]=(-1)^{mn}[v,u]$
\item[\textbf{(sa2)}] $[u,v\wedge w]=[u,v]\wedge w+(-1)^{(m+1)n}v\wedge[u,w]$
\item[\textbf{(sa3)}] $(-1)^{mk}[[u,v],w]+(-1)^{mn}[[v,w],u]+(-1)^{nk}[[w,u],v]=0$
\end{list}
The first of these three conditions is obviously always true. The
condition \textbf{(sa3)} is also always true for the the first
order elements and it can be easily seen that is equivalent to the
equality \[(\cd\circ\cd)(x\wedge y\wedge z)=0\] for any
$\{x,y,z\}\subset E_1$. This condition is equivalent to the
bracket $[x,y]=-\cd(x,y)$ be a Lie algebra structure on the space
$E_1$.\\
The condition \textbf{(sa2)} is equivalent to the following
equality for the operator \cd:
\begin{equation}
\begin{array}{l}
\cd(\alpha\wedge\beta\wedge\gamma)=\\
\\
=\cd(\alpha\wedge\beta)\wedge\gamma+(-1)^m\alpha\wedge\partial(\beta\wedge\gamma)+(-1)^{(m+1)n}\beta\wedge\partial(\alpha\wedge\gamma )-\\
\\
-(\partial\alpha\wedge\beta\wedge\gamma+(-1)^m\alpha\wedge\partial\beta\wedge\gamma+(-1)^{m+n}\alpha\wedge\beta\wedge\partial\gamma)
\end{array}
\end{equation}
It is easy to check that the condition \textbf{(sa2)} implies that
the operator \cd, on the elements of the type
$u_1\wedge,\ldots,\wedge u_n\in E_n$ where $u_1,\ldots,u_n$ are
elements of the space $E_1$, has the following form:
$$
\begin{array}{l}
\\{}
\cd(u_1\wedge,\ldots,\wedge u_n)=\\%
\\
=\sum_{i<j}(-1)^{i+j}[u_i,u_j]\wedge u_1\wedge\cdots\wedge\widehat{u_i}\wedge\cdots\wedge\widehat{u_j}\wedge\cdots\wedge u_n%
\\{}
\end{array}
$$
and in this case, all of the above three conditions are true
on the subalgebra $\wedge E_1=\sum_{k=0}^{\infty}(\wedge^kE_1)$.

Let $L$ is a Lie algebra and a module over the commutative algebra
$A$, which itself is a module over the Lie algebra $L$. That is:
there is a Lie algebra homomorphism from the Lie algebra $L$ to
the Lie algebra of derivations of the algebra $A$, and these two
structures are related via the following condition: for any
$x,y\in L$ and $a\in A$, we have $[x,ay]=x(a)\cdot y+a\cdot[x,y]$.

For any integer $n>0$, we denote by $\alt[n]{L,A}$ the space of
all skew-symmetric, multilinear (over the field of real or complex
numbers) mappings from $L^n$ to $A$. Using the
formula~\ref{CommutatorAsDeviation.Formula} for the
supercommutator on the space $\wedge L=\sum\wedge^kL$, we obtain
that for any $u\in\wedge^mL,\;v\in\wedge^nL$ and
$\omega\in\alt[m+n-1]{L,A}$ the following equality is true
\begin{displaymath}
\omega([u,v])=\omega(\cd(u){\wedge}~v)+(-1)^m\omega(u\wedge\cd(v))-\omega(\cd(u{\wedge}v))
\end{displaymath}
or in other notations
\begin{equation}\label{SupercommutatorInForm.Formula}
\begin{array}{l}
\omega([u,v])=\\
\\
=(-1)^{(m+1)n}i_v\omega(\cd(u))+(-1)^mi_u\omega(\cd(v))-\omega(\cd(u{\wedge}v))
\end{array}
\end{equation}
where, for $\alpha\in\alt[p]{L,A}$ and $x\in\wedge^qL$, under the
expression $i_x\alpha$, is denoted the element of the space
$\alt[p-q]{L,A}$ defined as \[(i_x\alpha)(y)=\alpha(x\wedge y)\]
for any $y\in\wedge^{p-q}L$.

Using the dual notations, the
expression~\ref{SupercommutatorInForm.Formula} can be rewritten in
the following form
\begin{equation}\label{SupercommutatorInFormStar.Formula}
\begin{array}{l}
\omega([u,v])=\\
\\
=(-1)^{(m+1)n}(\cd^{*}i_v\omega)(u)+(-1)^m(\cd^{*}i_u\omega)(v)-(\cd^{*}\omega)(u{\wedge}~v)
\end{array}
\end{equation}
where: $(\cd^{*}\alpha)(x)=\alpha(\cd(x))$, for any
$\alpha\in\alt[p]{L,A}$ and $x\in\wedge^pL$.

By the definition of the classical external differential, we have
that $d\alpha =\cd^{*}\alpha+\cd[1]\alpha$, where, for
$u_1,\ldots,u_{p+1}\in~{L}$, the expression $\cd[1]\alpha$ is
defined as
$$
(\cd[1]\alpha)(u_1\wedge\ldots\wedge~{u_{p+1}})=\sum_{i=1}^{p+1}(-1)^{i-1}u_i\alpha(u_1\wedge\ldots\wedge\widehat{u_i}\wedge\ldots\wedge~{u_{p+1}})
$$
It is easy to verify, that
$$
(-1)^{(m+1)n}(\cd[1]i_v\omega)(u)+(-1)^m(\cd[1]i_u\omega)(v)-(\cd[1]\omega)(u~{\wedge}~v)=0
$$
Therefore, in the expression
\ref{SupercommutatorInFormStar.Formula}, we can replace the
operator $\cd^{*}$ by the operator $d$, after which we obtain the
following formula for the Schouten bracket:
\begin{equation}\label{SchoutenBracketClassical.Formula}
\begin{array}{l}
\\{}
\omega([u,v])=\\
\\
=(-1)^{(|u|+1)|v|}(di_v\omega)(u)+(-1)^{|u|}(di_u\omega)(v)-(d\omega)(u~{\wedge}~v)
\\{}
\end{array}
\end{equation}
where $|u|$ denotes the order of the element $u\in \alt{L,A}$;
i.e., if $u$ is an element of the space $\alt[n]{L,A}$, then
$|u|=n$. This formula can be considered as an invariant definition
of the Schouten bracket in some cases, for example, in the case of
covariant, skew-symmetric tensor fields on a smooth manifold.
\subsection{Star Operator for a Poisson Structure. Poisson Cohomologies.}\label{Star}
As it was mentioned several times, an involutive element $p\in
L\wedge L$, defines the operator of degree $+1$
$$
\cd[p]=[p\,,\,\cdot]:\wedge L\To\wedge L
$$
which is a coboundary operator. The dual operator
$$
\cd[p]^{*}:\alt{L,A}\To\alt{L,A}
$$
defined as
$$
\cd[p]^{*}\omega)(x)=\omega([p,x]
$$
is a boundary operator ($\partial _p^{*}\circ\partial _p^{*}=0$)
of degree $-1$. Using the formula
\ref{SchoutenBracketClassical.Formula} we obtain the following
expression for $\cd[p]^{*}$:
$$
(\cd[p]^{*}\omega)(u)=(d\omega)(p\wedge u)-(di_p\omega)(u)-(-1)^{|u|}(di_u\omega)(p)%
$$
or, in more brief notations
\begin{equation}\label{dstarOperator.Formula}
(\cd[p]^{*}\omega)(u)=(i_p\circ d-d\circ i_p)(\omega)(u)-(-1)^{|u|}(di_u\omega)(p)%
\end{equation}
For any integer $n>1$, let $\xalt[n]{L,A}$ be the subspace of the
space $\alt[n]{A}$, consisting of the $A$-multilinear mappings.

It is clear that the subalgebra
$\xalt{L,A}=\sum\limits_{n=0}^{\infty}\xalt[n]{L,A}$
of the algebra $\alt{L,A}$ is not invariant under the action of
the operator $\cd[p]^{*}$, because, for
$\omega\in\xalt[n]{L,A},\;a\in A$, and $x\in\wedge^nL$, we have
$$
\begin{array}{l}
\\{}
(\cd[p]^{*}\omega)(a\cdot x)=-\omega([p,a\cdot x])=-\omega(\widetilde{p}(a)\wedge x+a[p,x])=\\
\\
=-((-1)^{|x|}(i_x\omega)(\widetilde{p}(a))+a\cdot\omega([p,x]))=\\
\\
=a\cdot(\cd[p]^{*}\omega)(x)-(-1)^{|x|}(i_x\omega)(\widetilde{p}(a))
\\{}
\end{array}
$$
To ``correct'' the operator $\cd[p]^{*}$, so that the algebra of
differential forms $\xalt{L,A}$ be invariant under its action, we
remove the last term in the \ref{dstarOperator.Formula}. The
result is exactly the boundary operator of the canonical complex
for Poisson manifold, which is well-known in the case when $L$ is
the Lie algebra of vector fields on some Poisson manifold $M$, and
$A$ is the commutative algebra of smooth functions on $M$
(see~\cite{brylinski}):
\begin{equation}\label{PoissonBoundaryOperator.Formula}
\delta:\xalt[m]{L,A}\To\xalt[m-1]{L,A},\quad\delta=i_p\circ\ud-\ud\circ i_p%
\end{equation}
Consider the following bilinear mapping
$$
\op{p}:\xalt[m]{L,A}\times\xalt[n]{L,A}\To\xalt[m+n-2]{L,A}%
$$
defined as
\begin{equation}\label{BivectorInWedge.Formula}
\op{p}(\alpha,\beta)=i_p(\alpha\wedge\beta)-i_p\alpha\wedge\beta-\alpha\wedge i_p\beta%
\end{equation}
The Schouten bracket on the anticommutative graded algebra
$$
\xalt{L,A}=\sum\limits_{k=0}^{\infty}\xalt[k]{L,A}
$$
can be defined as
\begin{equation}\label{SchoutenBracketByInWedge.Formula}
[~\alpha~,~\beta~]=\ud p(\alpha,\beta)-p(\ud\alpha ,\beta)-(-1)^{|\alpha|}p(\alpha,\ud\beta)%
\end{equation}
(see \cite{karasev-maslov}).
\begin{prop}
The bracket on $\xalt{L,A}$ defined by the formula
\ref{SchoutenBracketByInWedge.Formula} coincides with the bracket
$[~\cdot~,~\cdot~]_\delta$ which is the obstruction for the
operator $\delta$ to be an antiderivative. That is: for any
$\alpha\in\xalt[m]{L,A}$, and $\beta\in\xalt{L,A}$, the following
equality is true
$$
\delta\alpha\wedge\beta+(-1)^m\alpha\wedge\beta-\delta(\alpha\wedge\beta)=\ud p(\alpha,\beta)-p(\ud\alpha,\beta)-(-1)^mp(\alpha,\ud\beta)%
$$
\end{prop}
The proof of this proposition consists of a simple verifying of
the equation keeping in mind the formulas
\ref{PoissonBoundaryOperator.Formula},~\ref{BivectorInWedge.Formula}
and~\ref{SchoutenBracketByInWedge.Formula}.

For any $a\in A$, we the 1-form $\ud a\in\xalt[1]{L,A}$, as
$$
(\ud a)(X)=X(a)
$$
for any $X\in L$. Consider the subalgebra of the algebra
$\xalt{L,A}$ generated by $A$ and $\ud A\subset\xalt[1]{L,A}$. For
simplicity, further, we shall identify this subspace with the
entire space $\xalt{L,A}$. As it follows from the definition, this
space consists of the elements of the type
$$
\sum\limits_{i=1}^n a_0^i~\ud a_1^i\wedge\ud a_2^i%
$$
The Poisson bracket on $A$ defined by the involutive element $p$,
as $\{a,b\}=i_p(\ud a\wedge\ud b)$, for $a,b\in A$, gives the same
expression for the operator $\delta$, as in the case when $A$ is
the algebra of smooth functions on some Poisson manifold and $L$
is the Lie algebra of the vector fields on the same manifold
(see~\cite{brylinski}):
\begin{equation}\label{PoissonCoboundaryInCoordinates.Formula}
\begin{array}{l}
\\{}
\delta(a_0\ud a_1\wedge\ldots\wedge a_n)=\\
\\
=\sum\limits_{i=1}^n(-1)^{i+1}\{a_0,a_i\}\ud a_1\wedge\widehat{\ud a_i}\wedge \ud a_n+\\%
\\
+\sum\limits_{i<j}(-1)^{i+j}a_0\ud\{a_i,a_j\}\wedge\ud a_1\wedge\ldots\wedge\widehat{\ud a_i}\wedge\ldots\wedge\widehat{\ud a_j}\wedge\ldots\wedge\ud a_n%
\end{array}
\end{equation}
By using of this formula, it is easy to verify that on
$\xalt{L,A}$ the condition \ref{SchoutenBracketByInWedge.Formula}
for $\delta$ is true. Therefore, the bracket defined by
\ref{PoissonBoundaryOperator.Formula} or by
$$
[\alpha,\beta]=\delta\alpha\wedge\beta+(-1)^{|\alpha|}\alpha\wedge\delta\beta-\delta(\alpha\wedge\beta)%
$$
on $\xalt{L,A}$ gives a \emph{Lie superalgebra structure}, which
is the extension of the Lie algebra structure on the space
$\xalt[1]{L,A}$. So, in the case when $A=C^\infty(M)$ for some
Poisson manifold $M$, and $L$ is the Lie algebra of vector fields
on the same manifold, we can state that the supercommutator of
differential forms on $M$ is the obstruction for the canonical
boundary operator $\delta$ to be an antidifferential.

An element $\xi=x\wedge y\in L\wedge L$ defines the mapping from
$A$ into $L$, $a\mapsto\bop{\xi}{a}$, as
$$
\bop{\xi}{a}=x(a)\cdot y-y(a)\cdot x
$$
It is clear that for each $a\in A$, the expression $\bop{\xi}{a}$
depends only on $\ud a\in\xalt[1]{L,A}$. This mapping can be
extended linearly for any $p\in L\wedge L$. After that, for any
$p\in L\wedge L$ we can define the mappings
$$
\widetilde{p}:\xalt[k]{L,A}\To\wedge^kL,\quad k=0,\ldots,\infty
$$
as follows
$$
\bop{p}{\,{a_0\,\ud a_1\wedge\cdots\wedge\ud a_k}}=a_0\,\bop{p}{a_1}\wedge\cdots\wedge\bop{p}{a_k}%
$$
For any fixed $\omega\in\xalt[n]{L,A}$, define a series of
mappings:
$$\star:\xalt[k]{L,A}\To\xalt[n-k]{L,A}$$
as
$$
\star\,(a_0\,\ud a_1\wedge\ldots\wedge\ud
a_k)=a_0\,(i_{\bop{p}{a_k}}\circ\cdots\circ i_{\bop{p}{a_1}})(\,\omega\,)%
$$
In the case when $M$ is a \textbf{symplectic} manifold with a
\textbf{symplectic} form $\alpha$, the commutative algebra $A$ is
$C^\infty (M)$, the second order element $p$ is the
\textbf{bivector} field corresponding to the form $\alpha$, the
Lie algebra $L$ is the Lie algebra of vector fields on $M$, and
$\omega=\alpha^{(\dim M)/2}$, the operator $\star$ is the
well-known analogue of the star operator on a \textbf{Riemannian}
manifold (see~\cite{hodge},~\cite{warner},~\cite{brylinski}).
\begin{prop}
If $\omega$ satisfies the following conditions $d\omega=0$ and
$\ud a\wedge\omega=0$ for each $a\in A$ then the equality
$\star\circ\delta =(-1)^k\ud\circ\star$ is true on the algebra
$\xalt[k]{L,A}$, if and only if $(d\circ i_{\bop{p}{a}})\omega=0$,
for any $a\in A$.
\end{prop}
\textbf{Proof.} On the space $\xalt[1]{L,A}$ we have the following
equalities:
$$%
(\star\circ\delta)(a_0\ud a_1)~=~\star(\{a_0,a_1\})~=~\{a_0,a_1\}\cdot\omega%
$$%
and
$$%
(\ud\circ\star)(a_0\,\ud a_1)=\ud(a_0\,i_{\bop{p}{a_1}}\omega)=\ud a_0\,\wedge i_{\bop{p}{a_1}}\omega+a_0\,\ud i_{\bop{p}{a_1}}\omega%
$$%
Consequently:
$$
\begin{array}{l}
(\star\circ\delta+\ud\circ\star)(a_0\,\ud a_1)=\\
\\
=\{a_0,a_1\}\cdot\omega+\ud a_0\wedge i_{\bop{p}{a_1}}\omega+a_0\,\ud i_{\bop{p}{a_1}}\omega=\\%
\\
=-i_{\bop{p}{a_1}}(\ud a_0\wedge\omega)+a_0\,\ud i_{\bop{p}{a_1}}\omega=a_0\,\ud i_{\bop{p}{a_1}}\omega%
\end{array}
$$
Therefore, on the space $\xalt[1]{L,A}$ the equality
$\star\circ\delta =-\ud\circ\star$ is true if and only if
$(d\circ i_{\bop{p}{a}})\omega=0$
for any $a\in A$.

To proof the equality $$\star\circ\delta =(-1)^k\ud\circ\star$$
for every space $\xalt[k]{L,A},\;k=0,\ldots,\infty$, the following
well-known formula can be used
$$
\begin{array}{l}
\\{}
(L_X\omega)(X_1,\ldots,X_n)=\\
\\
=(i_X\ud\omega+\ud i_X\omega)(X_1,\ldots,X_n)=\\%
\\
=X\omega(X_1,\ldots,X_n)-\sum\limits_i\omega(X_1,\ldots,[X,X_i],\ldots,X_n)
\end{array}
$$
$\Box$

So, we can state that the operator $\star$ induces a homomorphism
from the homology space $H_i(L,\,A,\,\delta)$ of the complex
$\big(\xalt{L,A},\;\delta\big)$ into the cohomology space
$H^{n-i}(L,\,A)$ of the complex $\big(\xalt{L,A},\;\ud\big)$.
\newpage
\section{Differential Complex and Generalized Functions on Poisson Manifold}
\subsection{Brief Overview of Geometric structures on Poisson
Manifold} Further we shall consider the case when the commutative
algebra $A$ is the algebra $C^\infty(M)$ for some \smooth class
manifold $M$; the Lie algebra $L$ is the Lie algebra of vector
fields on the manifold $M$ and therefore $\xalt{L,A}$ is the
exterior algebra of differential forms on $M$, denoted here by
$\Omega(M)$. An involutive element $p\in V^2(M)$, where $ V^2(M)$
is the space of the second-order covariant antisymmetric tensor
fields on $M$, defining a Poisson algebra structure on the space
$C^\infty (M)$ is called as a \emph{bivector field} on the
manifold $M$, correspondent to the Poisson structure. For the
Poisson bracket of a pair of smooth functions on $M$, we use the
common notation: $\{f,\,g\}$.

In the case when the bivector field $p$ is a non-degenerated
covariant tensor field, it defines a second order differential
form $\omega$ on the manifold $M$, which is non-degenerated and
the condition $[p\,,\,p]=0$ is equivalent to $\der{\omega}=0$. In
this case, the differential form $\omega$ is called the
\emph{symplectic form} and the pair $(M,\,\omega)$ is said to be a
\emph{symplectic manifold}. A vector field $X\in\vecf{M}$ is said
to be \emph{symplectic}, if $L_X(\omega)=0$. For a symplectic
manifold $(M,\,\omega)$, let us denote by $\vecf{M}_S$ the space
of all symplectic vector fields on this manifold. From the
following well-known equalty $L_{[X,\,Y]}=[L_X,\,L_Y]$, easily
follows that the space $\vecf{M}_S$ is a Lie subalgebra in
$\vecf{M}$, though it is not a $\smooth{(M)}$-submodule.

In the case of a symplectic manifold, the mapping
$\widetilde{p}:\Omega^1(M)\To\vecf{M}$, is an isomorphism, and its
restriction on the subspace of the closed 1-forms, it gives an
isomorphism to the space symplectic vector fields on $M$. This
isomorphism allows us to carry the Lie algebra structure from
$\vecf{M}_S$ to the space $Z^1(M)$ --- the space of closed
one-form on the manifold $M$. It is clear, that
$$
i_{[|X,\,Y]}\omega=L_X\big(i_Y\omega\big)=\der{\big(\big(i_X\circ i_Y\big)(\omega)\big)}
$$
which implies that $B^1(M)$ --- the space of exact 1-forms on the
manifold $M$ is a Lie algebra ideal in $Z^1(M)$. Moreover:
$$[Z^1(M),\,Z^1(M)]\subset B^1(M)$$
(see~\cite{hurt}).

Let us denote by $\vecf{M}_H$ the subalgebra of the Hamiltonian
vector fields on the manifold $M$. It is clear that
$\vecf{M}_H\subset\vecf{M}_S$. From $[Z^1(M),\,Z^1(M)]\subset
B^1(M)$ follows that
$$\big[\vecf{M}_S,\,\vecf{M}_S\big]\subset\vecf{M}_H$$
Let $G$ is a Lie group that acts on the manifold $M$, so that for
any $g\in G$, the corresponding diffeomorphism $g:M\To M$ is a
symplectic one: $g^*(\omega)=\omega$. In this case, the triple
$(M,\,\omega,\,G)$ is said to be a \emph{symplectic $G$-space}. If
the action of the group $G$ is transitive, then it is called the
\emph{homogeneous symplectic $G$-space}.

For a symplectic $G$-space $(M,\,\omega,\,G)$, we have the
canonical mapping $\sigma:\lie{g}\To\vecf{M}_S$, which assigns to
each element $u$ of the Lie algebra of the Lie group $G$, the
vector field $\sigma(u)$ corresponding to the one parameter group
of diffeomorphisms $\exp(t\cdot u)$.

A symplectic $G$-space $(M,\,\omega,\,G)$ is said to be
\emph{strictly symplectic} if for for any
$u\in\lie{g}:\;\sigma{u}\in\vecf{M}_H$.

Let $h:C^{\infty}(M)\To\vecf{M}$ be the \emph{Hamiltonian}
mapping, which assigns to each smooth function $f\in
C^{\infty}(M)$ the corresponding Hamiltonian vector field $X_f$. A
mapping $\lambda:\lie{g}\To C^{\infty}(M)$ is called the
\emph{momentum map}, if $h\circ\lambda=\sigma$.

Let us consider the following exact sequence
$$
0\To{}H^0(M,\,\mathbf{R})\stackrel{\mu}{\To}C^{\infty}(M)\stackrel{h}{\To}\vecf{M}_S\stackrel{\eta}{\To}H^1(M,\,\mathbf{R})\To{}0
$$
The canonical Lie algebra structure on the space $H^1(M)$ is
trivial, and from
$\big[\vecf{M}_S,\,\vecf{M}_S\big]\subset{}h\big(C^{\infty}(M)\big)$,
follows that the mapping $\eta$ is a Lie algebra homomorphism. In
the same manner, the mapping $\mu$ is a Lie algebra homomorphism
too. Therefore, it can be stated that the above exact sequence is
an exact sequence of \textbf{Lie algebra homomorphisms}. A
momentum map $\lambda:\lie{g}\To C^{\infty}(M)$ exists iff
$\textbf{Image}(\sigma)\subset h(C^{\infty}(M))$, which is
equivalent $\eta\circ\sigma=0$. As the composition mapping
$\eta\circ\sigma$ is a homomorphism of Lie algebras, and the Lie
algebra $H^1(M)$ is commutative, we have that
$(\eta\circ\sigma)([\lie{g},\lie{g}])=0$. Therefore, the mapping
$\eta\circ\sigma\equiv\theta$ induces the quotient mapping
$$\widetilde{\theta}:\lie{g}/[\lie{g},\lie{g}]\To H^1(M)$$
After this, it can be stated that, the necessary and sufficient
condition, for existence of the momentum mapping is the equality
$\widetilde{\theta}=0$. This condition can be satisfied in the
following three cases (see~\cite{hurt}):
\begin{list}{}{\topsep=10pt\itemsep=10pt}
\item[\textbf{Case 1:}]
the Lie algebra $\lie{g}$ is semisimple
($[\lie{g},\,\lie{g}]=\lie{g}$);
\item[\textbf{Case 2:}]
if the differential form $\omega$ is exact: $\omega=\der{\alpha}$
and the 1-form $\alpha$ is invariant under the action of the Lie
group $G$: for any $u\in\lie{g}$, $L_{\sigma(u)}\alpha=0$; in this
case, we have an explicitly defined momentum mapping
$\lambda(u)=-\alpha(\sigma(u))$;
\item[\textbf{Case 3:}]
$H^1(M)$ -- the first De Rham cohomology space of the manifold
$M$, is trivial one.
\end{list}

To investigate the question: \textbf{is the momentum map
$$\lambda:\lie{g}\To C^{\infty}(M)$$ a Lie algebra homomorphism or
not?} consider the following construction:\\
let the manifold $M$ be strictly symplectic (i.e,
$\textbf{image}(\sigma)\subset\vecf{M}_H$). Consider the mapping
$m:M\To\lie{g}^*$, defined as
$$
m(x)(u)\;=\;\lambda(u)(x)
$$
for $x\in M$ and $u\in\lie{g}$; and a bilinear mapping
$$
c\,:\,\lie{g}\,\times\,\lie{g}\;\To\;C^{\infty}(M)
$$
defined as
$$
c(u_1\,,\,u_2)\;=\;m\big([u_1\,,\,u_2]\big)-\big\{m(u_1)\,,\,m(u_2)\big\}
$$
The mapping $c$ is obviously antisymmetric; the composition
mapping $h\circ c$ is trivial and for any $u,v,w\in\lie{g}$ we
have that
$$
c(\,[\,u\,,\,v\,]\,,\,w\,)+c(\,[\,v\,,\,w\,]\,,\,u\,)+c(\,[\,w\,,\,u\,]\,,\,v\,)\;=\;0
$$
The latter equality, together with the antisymmetricity is the
condition for the mapping $c$, being a two-dimensional cocycle for
the Lie algebra $\lie{g}$, with values in the ring
$C^{\infty}(M)$.\\
\textbf{The momentum map $\lambda:\lie{g}\To C^{\infty}(M)$ is a
Lie algebra homomorphism if and only if the above defined
2-dimensional cocycle $c$, is trivial} (see~\cite{hurt}).

Now, let us return to the general situation, when the bivector
field $p$ on a Poisson manifold $M$, is not necessarily
non-degenerated (i.e., the manifold $M$ is not necessarily
symplectic one).

Let $\pi$ be the differential system on $M$ derived by the set of
the vector fields of the type $X_f=\{f,\,\cdot\}$ for $f\in C^\infty(M)$.%
In other words, for any point $x\in M$, the subspace $\pi(x)\subset\tang[x]{M}$%
is defined as
$$
\pi(x)=\{\,u_{\alpha}\in\tang[x]{M}\;|\;\beta(u_{\alpha})=(\alpha\wedge\beta)\big(p(x)\big)\,:\,\alpha\,,\,\beta\in\ctang[x]{M}\,\}
$$
The \textbf{rank} of a bivector field at any point $x\in M$ is
defined as
$$
\rk{p}=2k \iff \wedge^kp(x)\neq 0 \textrm{ and } \wedge^{k+1}p(x)\,=\,0%
$$
For any function $f\in C^\infty (M)$, let
$$
\{\,\Phi^f_t\;|\;t\in\mathbf{R}\,\}
$$
be the one-parameter group of diffeomorphisms of the manifold $M$,
corresponding to the vector field $X_f=\{\,f\,,\,\cdot\}$.

For any two functions $g,h\in C^{\infty}(M)$ and a point $x\in M$,
consider the function $\lambda_x:\mathbf{R}\To\mathbf{R}$ defined
as
$$\lambda_x(t)=\big(\{g\circ\Phi^f_t,h\circ\Phi^f_t\}-\{g,h\}\circ\Phi^f_t\big)(x)$$
As $\Phi^f_t\,,\,t\in \mathbf{R}$ is a one-parameter group of
diffeomorphisms corresponding to some vector field, the element
$\Phi^f_t$ is the identical mapping. Therefore, we have that
$\lambda_x(0)=0$. Also, it is clear that
$$
\dot{\lambda_x}(t)=\big(\{\,X_f(g)\,,\,h\,\}+\{\,g\,,\,X_f(h)\,\}-X_f(\{\,g\,,\,h\,\})\big)\big(\Phi^f_t(x)\big)
$$
which is equivalent to
$$
\dot{\lambda_x}(t)=\big(\{\{f,g\},h\}+\{g,\{f,h\}\}-\{f,\{g\,,\,h\}\big)\big(\Phi^f_t(x)\big)
$$
Take into consideration the Jacoby identity for the Poisson
bracket $\{\,\cdot\,,\,\cdot\}$, the last equality implies that
$\dot{\lambda_x}(t)=0$ for any $x\in M$. Hence, we obtain that
$$
\{g\circ\Phi^f_t,h\circ\Phi^f_t\}=\{g,h\}\circ\Phi^f_t
$$
which is equivalent to
$$
\big(\ud(g\circ\Phi^f_t)\wedge\ud(h\circ\Phi^f_t)\big)(p)\circ(\Phi^f_t)^{-1}=(\ud g\wedge\ud h)(p)%
$$
The last equality implies that the bivector field $p$ is invariant
under the action of the one-parameter group $\Phi^f_t$ for every
smooth function $f\in C^{\infty}(M)$.

It is natural to ask: \textbf{is the differential system $\pi$
integrable or not?}

Note that $\pi$ is an involutive differential system:
$$
\begin{array}{l}
(X,Y)\in\pi\Leftrightarrow\big(X=\sum\limits_i\varphi_i\{\,f_i\,,\,\cdot\,\},\;Y=\sum\limits_i\psi_i\{\,g_i\,,\,\cdot\,\}\big)\Rightarrow\\%
\Rightarrow[\,X\,,\,Y\,]=\sum\limits_i\big(\varphi_i\{\,f_i\,,\,\psi_i\,\}\cdot\{\,g_i\,,\,\cdot\,\}-\psi_i\{\,g_i\,,\,\varphi_i\,\}\{\,f_i\,,\,\cdot\,\}\big)+\\%
+\sum\limits_i\varphi_i\psi_i\,\{\,\{\,f_i\,,\,g_i\,\}\,,\,\cdot\,\}\in\pi
\end{array}
$$
Moreover, the following theorem gives the exact condition for any
bivector field $p$, the corresponding differential system be
involutive:
\begin{thm}
The differential system $\pi$ corresponding to a bivector field
$p$ is involutive \textbf{if and only if}
$$[\,p\,,\,p\,](x)\in\pi(x)\wedge\pi(x)\wedge\pi(x)$$ for every
point $x\in M$.
\end{thm}
\textbf{Proof.} For proof, the following formula is useful: for
$\omega\in\Omega^2(M),\;\alpha,\beta\in\Omega^1(M)$ and $X,Y\in V^2(M)$%
\begin{equation}\label{Formula14}
\begin{array}{l}
(\omega\wedge\alpha\wedge\beta)(X\wedge Y)=\\%
\\
=\omega(X)\cdot(\alpha\wedge\beta)(Y)+\omega(Y)\cdot(\alpha\wedge\beta)(X)-\\%
\\
-\omega(\widetilde{X}(\alpha),\widetilde{Y}(\beta))+\omega(\widetilde{X}(\beta),\widetilde{Y}(\alpha))
\end{array}
\end{equation}
Where $\widetilde{X}$ and $\widetilde{Y}$ are the mappings from
$\Omega^1(M)$ to $V^1(M)$ defined by the
formula~\ref{tlldeBivectorFieldMapping.Formula}. It is sufficient
to verify this formula in the case when
$\omega=\varphi\wedge\psi$, for any $\varphi,\psi\in\Omega^1(M)$.
In this case we have the following:
$$
\begin{array}{l}
\\{}
(\varphi\wedge\psi\wedge\alpha\wedge\beta)(X\wedge Y)=(\varphi%
\wedge\psi)(X)\cdot(\alpha\wedge\beta)(Y)+\\
\\
+(\varphi\wedge\alpha)(X)\cdot(\beta\wedge\psi)(Y)+(\varphi\wedge%
\beta)(X)\cdot(\varphi\wedge\alpha)(Y)+\\
\\
+(\psi\wedge\alpha)(X)\cdot(\varphi\wedge\beta)(Y)+(\psi\wedge%
\beta)(X)\cdot(\alpha\wedge\varphi)(Y)+\\
\\
+(\alpha\wedge\beta)(X)\cdot(\varphi\wedge\psi)(Y)=\\
\\
=\omega(X)\cdot(\alpha\wedge\beta)(Y)+\omega(Y)\cdot(\alpha\wedge%
\beta)(X)-\\
\\
-\omega(\widetilde{X}(\alpha),\widetilde{Y}(\beta))+%
\omega(\widetilde{X}(\beta),\widetilde{Y}(\alpha))
\\{}
\end{array}
$$
The statement of the theorem, translated on the language of a
local coordinate system $\{x_1,\ldots,x_n\}$ is the following: for
each $i,j\in\{1,\ldots,n\}$ the vector field
$[\,\widetilde{p}(dx_i)\,,\,\widetilde{p}(dx_j)\,]$ takes its
values in the differential system $\pi$; which is the same as,
that
$\sigma([\,\widetilde{p}(dx_i)\,,\,\widetilde{p}(dx_j)\,])=0$
for each $\sigma\in(\pi)^{\perp}\subset\Omega^1(M)$.\\
Using the formula \ref{SchoutenBracketClassical.Formula} for the
Schouten bracket, we obtain:
$$
\begin{array}{l}
\big(\ud\sigma\wedge \ud x_i\wedge\ud x_j\big)(p\wedge p)=\\
\\
=(2\ud\sigma)(p)\cdot(\ud x_i\wedge\ud x_j)(p)-\big(\sigma\wedge\ud x_i\wedge \ud x_j\big)([p,p])%
\\{}
\end{array}
$$
By using of the formula \ref{Formula14}, we obtain:
$$
\begin{array}{l}
\\{}
\big(d\sigma\wedge\ud x_i\wedge\ud x_j\big)(p\wedge p)=\\
\\
=(2\ud\sigma)(p)\cdot(\ud x_i\wedge\ud x_j)(p)-\big(2\sigma\big)\big(\widetilde{p}(\ud x_i),\widetilde{p}(\ud x_j)\big)%
\\{}
\end{array}
$$
and finally, we have the following:
$$
\big(\sigma\wedge\ud x_i\wedge\ud x_j\big)([p,p])=-(2\sigma)\big(\widetilde{p}(\ud x_i),\widetilde{p}(\ud x_j)\big)%
$$
take into consideration the fact that $\sigma\in(\pi)^{\perp}$,
the last equality ends the proof of the theorem.
$\Box$

If the rank of a differential system $\pi$ (or, which is the same,
the rank of the tensor field $p$) is constant, then its
integrability follows from the Frobenius's classical theorem; but
generally, the differential system $\pi$, is not of a constant
rank. Recall, that as it follows from the Hermann's theorem, the
necessary and sufficient condition for the integrability of a
differential system is the conservation of the rank of the system
along the integral paths of this system. This condition is
satisfied for the differential system $\pi$, and it follows from
the fact that the bivector field $p$ is invariant under the action
of one-parameter group of the Hamiltonian vector field
corresponding to any smooth function on the manifold.

An integral leaf of the differential system $\pi$ is called a
\emph{symplectic leaf} of Poisson structure $p$.

The restriction of the Poisson structure $p$ on any integral leaf
of the differential system $\pi$ is non-singular because the rank
of this restriction coincides with the dimension of the leaf;
hence, such restriction induces a symplectic structure on this
leaf.

Let us denote the symplectic form on a symplectic leaf $N$
corresponding to the restriction of the Poisson structure on this
leaf by $\omega_N$. For $x\in N,\;u\in\tang[x]{N}$, and
$v\in\tang[x]{N}$ we have that $\omega_N(u,v)~=~\{f,g\}(x)$, where
$f$ and $g$ are such functions on the manifold $M$ that
$u=\{\,f\,,\,\cdot\,\}(x)$, and $v=\{\,g\,,\,\cdot\,\}(x)$.

One of the indicators of the singularity of a Poisson structure is
the existence of such non-constant smooth function on $M$, which
commutes with all smooth functions on $M$, i.e. the center of the
Lie algebra of smooth functions, does not coincide to the set of
the constant functions. The elements of the center $Z(M)$ are
referred as \textbf{Casimir} functions. From the singularity of
the Poisson structure $p$ does not follow the existence of a
non-constant Casimir function. For instance, if one of the
symplectic leaves of the Poisson structure is everywhere dense in
the manifold $M$ then a Casimir function can be only constant
function. To illustrate this more explicitly, consider the
following 
\begin{example}\label{Example1}
let $M$ be a two-dimensional symplectic manifold and $p$ be the
corresponding non-singular bivector field on $M$. Let $\varphi $
be a non-constant smooth function on $M$. The bivector field
$p_1=\varphi\cdot p$ , is involutive as well as $p$. If the set
$\varphi^{-1}(0)$ is not empty, the Poisson structure defined by
$p_1$, is singular at the points of the set $\varphi^{-1}(0)$,
which follows from the relation between the bracket
$\{\,\cdot\,,\,\cdot\,\}_1$ defined by the bivector field $p_1$
and the bracket $\{\,\cdot\,,\,\cdot\,\}$ defined by the
symplectic structure $p$
$$
\{\,f\,,\,g\,\}_1=\varphi\cdot\{\,f\,,\,g\,\}
$$
for any $f,g\in C^\infty(M)$.

If a function $f\in C^\infty(M)$ is a Casimir function, then we
have the following
$$
\big(\,\varphi\cdot\{\,f\,,\,\cdot\,\}=0\,\big)%
\Rightarrow\big(\,\{\,f\,,\,\cdot\,\}\big|_{M\setminus\varphi^{-1}(0)}=0\,\big)%
\Rightarrow\big(\,f=\textbf{const}\,\big)%
$$
If $\varphi$ is such function, that the set
$M\setminus\varphi^{-1}(0)$ is everywhere dense set in the
manifold $M$ (for example, in the case when the set
$\varphi^{-1}(0)\}$ consists only one point $x_0$), then we have
that the function $f$ is constant on the manifold $M$. So, this is
an example of the situation when a Poisson structure is singular,
but Casimir function can be only constant.
\end{example}
Further, we shall extend (in some sense) the definition of the
Poisson bracket on distributions (\textbf{generalized functions})
on a smooth manifold, and be looking for Casimir functions in the
set of distributions.
\subsection{Brief Review of Distributions (Generalized Functions) on Smooth Manifold}
The theory of generalized functions is a common method which
allows to manipulate of divergent integrals and series, to
differentiate non-smooth functions and perform several such kind
of operations over various singular objects. In the following two
sections, we use the language of the generalized functions to
extend some algebraic notions common for regular Poisson manifolds
to the case of singular ones.

For a smooth manifold $M$, let $C_0^{\infty}(M)$ be the subalgebra
of the algebra $C^{\infty}(M)$, consisting of the functions with
compact support. Any linear, continuous functional on the space
$C_0^{\infty}(M)$ we call a \textbf{generalized function} or
\textbf{distribution} on the manifold $M$. The space of all
generalized functions on the manifold $M$ we denote further by
\dist{M}. Using the classical notation, the value of a generalized
function $\Phi$ on a smooth function with compact
support$\varphi$, we denote by
$\langle\,\Phi\,,\,\varphi\,\rangle$.

Let $M$ be a $n$-dimensional oriented manifold with volume form
$\vol\in\Omega^n(M)$. A function $f$, on the manifold $M$, is
called a \emph{locally integrable} function, if for any compact
subset $K$ of the manifold $M$ the restriction of the function $f$
on the subset $K$ is integrable under the volume form $\vol$. Such
function, defines a functional $[f]:\comp{M}\To\mathbf{R}$ via the
action: for any $\varphi\in\comp{M}$ let
$$
[f](\varphi)=\int\limits_Mf\varphi\cdot\vol
$$
This functional is evidently linear and smooth and subsequently is
a generalized function. If we denote by $F(M)$ the space of
locally integrable functions on the manifold $M$, it can be stated
that there is a mapping $F(M)\To\dist{M},\;f\mapsto[f]$. If two
functions $f,g\in F(M)$ differ only on a $0$-measure subset of the
manifold $M$, then we have that $[f]=[g]$; and inversely: for any
two functions $f,g\in F(M)$, if $[f]=[g]$, then they differ only
on a $0$-measure subset of the manifold $M$ (or in other words:
they are equal almost everywhere on the manifold $M$). The image
of the space $F(M)$ in \dist{M} is dense under the topology of
weak convergency. Moreover, any generalized function on the
manifold $M$ is a weak limit of the sequence of smooth functions
on the manifold $M$.

Let us review some algebraic and differential operations over the
space of generalized functions on the manifold $M$. These
operations are in concordance with the analogical operations on
the image of the space $F(M)$, i.e., the mapping $F(M)\To\dist{M}$
is a homomorphism under these operations. Here is the list of
these operations:
\begin{list}{}{\topsep=10pt\itemsep=10pt}
\item[\textbf{Addition: }]
For any two generalized functions $f,g\in\dist{M}$ and a smooth
function $\varphi\in\comp{M}$, let
$$\scalar{f+g}{\varphi}=\scalar{f}{\varphi}+\scalar{g}{\varphi}$$
\item[\textbf{Multiplication: }]
It can be defined a \textbf{multiplication of a generalized
function on a smooth function}. For $f\in\dist{M}$, $\phi\in
C^{\infty}(M)$ and $\psi\in\comp{M}$, let
$$
\scalar{\phi\cdot f}{\psi}=\scalar{f}{\phi\cdot\psi}
$$
This operation makes the space $\dist{M}$ a \textbf{module} over
the algebra $C^{\infty}(M)$.
\item[\textbf{Differentiation: }]
For any vector field $X\in V^1(M)$, a generalized function
$f\in\dist{M}$ and a smooth function $\varphi\in\comp{M}$, let
$$
\scalar{X(f)}{\varphi}=-\scalar{f}{X(\varphi)}
$$
This operation defines a \textbf{connection for the pair} (see
Definition~\ref{ConnectionForPair.Definition})
$\big(\,V^1(M)\,,\,\dist{M}\,\big)$. That is:\\
for any $f\in\dist{M},\;\varphi\in C^{\infty}(M)$ and $X\in
V^1(M)$, we have the following
$$
X(\varphi\cdot f)=X(\varphi)\cdot f+\varphi\cdot X(f)%
$$
This equality follows from the following series of equalities
$$
\begin{array}{l}
\\{}
\scalar{X(\varphi\cdot f)}{\psi}=-\scalar{\varphi\cdot f}{X(\psi)}=\\%
\\
=-\scalar{f}{\varphi\cdot X(\psi)}=\\%
\\
=\scalar{f}{X(\varphi)\cdot\psi}-\scalar{f}{X(\varphi\cdot\psi)}=\\%
\\
=\scalar{X(\varphi)\cdot f}{\psi}+\scalar{X(f)}{\varphi\cdot\psi}=\\%
\\
=\scalar{X(\varphi)\cdot f}{\psi}+\scalar{\varphi\cdot X(f)}{\psi}%
\\{}
\end{array}
$$
Let $U$ be an open subset of the manifold $M$. Let \comp{M,U} be
the subspace of the space \comp{M}, consisting of the smooth
functions $\varphi\in\comp{M}$ such that: $\varphi=0$ outside of
some compact set $K\subset U$. The restriction of a generalized
function $f\in\dist{M}$ on the open subset $U\in M$ is a
functional $$f|_U:\comp{M,U}\To\mathbf{R}$$ which is the
restriction of the functional $f:\comp{M}\To\mathbf{R}$ to the
subspace $\comp{M,U}$.
\item[\textbf{Push-forward: }]
Let $N$ be another smooth manifold and $$F~:~M~\To~N$$ be any
smooth mapping. For any generalized function $f\in\dist{M}$, let
$F_*(f)$ (push-forward) be a generalized function on the manifold
$N$, defined as
$$
F_*(f)\big(\varphi\big)=f\big(\varphi\circ F\big)
$$
for any $\varphi\in\comp{M}$.\\
Using this definition, we can define the right action of the group
of diffeomorphisms of the manifold $M$, on the space \dist{M}: for
any $G\in \textrm{Diff}(M)$ and $f\in \dist{M}$, let
$$
Gf=G_*^{-1}(f)
$$
\end{list}

A generalized function $f\in\dist{M}$ is said to be \emph{equal to
zero} on an open subset $U\subset M$, iff the restriction of $f$
on the subset $U$ equal to zero: $f|_U=0$. Let the open set
$V\subset M$ be the union of all open subset of $M$ on which the
generalized function $f$ equal to zero. The complement of the
subset $V$: $M\setminus V$, is called the \emph{support} of the
generalized function $f$, and is denoted by $\textbf{supp}(f)$.

Let a commutative algebra $A$ is equal to the algebra of smooth
functions on the manifold $M$, and $P=\Gamma(\pi)$ be a module of
smooth sections of some vector bundle over the manifold $M$. In
this case, the vector space (fiber) $\pi^{-1}(x)$ for any point
$x\in M$ is canonically isomorphic to the quotient module
$P_x=P/(I_x\cdot P)$, where $I_x$ is a module in the algebra $A$,
consisting of the functions equals to $0$, in the point $x$; and
$I_x\cdot P$ denotes the submodule of the module $P$, generated by
the elements of the type $\varphi\cdot s,\;\varphi\in
I_x,\,s\in\Gamma$. Via this isomorphism, the evaluation mapping
$P\ni s\mapsto s(x)\in\pi^{-1}(x)$, corresponds to the natural
quotient mapping $q_x:P\To P_x$. It is clear that if for some
element $s\in P,\;q_x(s)=0,\;\forall x\in M$, then the element $s$
equal to zero. In general algebraic situation, when $P$ is any
module over the algebra $A=C^{\infty}(M)$, it can happen that for
some element $s\in P$, we have that $q_x(s)=0$ for all points
$x\in M$ but the element $s$ is not equal to $0$. Such type of
$A$-module, cannot be realized as a module of sections of some
vector bundle. They are some kind of \textbf{non-geometric}
modules. Now, let us give more strict description of this
situation.

A family of elements $\{p_i\,|\,i\in I\}\subset P$ is called a
\emph{generated family} for the $A$-module $P$, if any element of
$P$ can be represented (possibly in more that one manner) as a sum
$\sum\limits_{i\in I}a_ip_i$, with $a_i\in A$, where only a finite
number of terms in the sum are different from zero. The family
$\{p_i\,|\,i\in I\}$ is called \emph{free} if it is made of
linearly (over the algebra $A$) independent elements, and it is a
\emph{basis} for the module $P$ if it is a \textbf{free generating
family}; that is: any $s\in P$ can be represented
\textbf{uniquely} as a linear combination $\sum\limits_{i\in
I}a_ip_i$. The module $P$ is called free if it admits a basis; and
is said to be of \emph{finite type} if it is \textbf{finitely
generated}, i.e., if it admits a generating family with finite
number of elements (see~\cite{landi}).
\begin{defn}[see~\cite{landi}]
A module over the algebra $A$ is said to be \emph{projective} if
it satisfies the following three equivalent properties:
\begin{list}{}{\leftmargin=10pt\topsep=10pt\itemsep=10pt}
\item[\textbf{(pr1)}]
For any epimorphism $\phi:P_1\To P_2$ of $A$-modules, any
homomorphism $f:P\To P_2$ can be lifted to a homomorphism
$$\tilde{f}~:~P~\To~P_1$$ such that $\phi\circ\tilde{f}=f$;
\item[\textbf{(pr2)}]
Every epimorphism $f:P_1\To P$ can be split; i.e., there exists a
homomorphism $s:P\To P_1$ such that $f\circ s=Id_P$;
\item[\textbf{(pr3)}]
The module $P$ is a direct summand of some free module; i.e.,
there exist a free module $\Gamma$ and a module $P'$, such that
$\Gamma=P\oplus P'$.
\end{list}
\end{defn}
The following central statement, provides the criteria for module
over the algebra of smooth functions on a smooth manifold to be a
module of sections of some vector bundle over this manifold
(see~\cite{swan}, \cite{connes4}, \cite{landi}).
\begin{prop}
Let $M$ be a compact finite dimensional manifold. A
$C^{\infty}(M)$-module $P$ is isomorphic to a module of smooth
sections of some vector bundle over the manifold $M$, if and only
if $P$ is a finite projective module.
\end{prop}

For any point $x\in M$, let $\delta_x\in\dist{M}$ be so-called
\textbf{Dirac function}; i.e., it is a linear functional
$\delta_x:\comp{M}\To\mathbf{R}$ defined as
$$\delta_x(\varphi)=\varphi(x)$$
for any $\varphi\in C^{\infty}(M)$. By definition, its derivation
is
$$\delta_x'(\varphi)=-\delta_x(\varphi')=-\varphi'(x)$$
\begin{lem}
For any point $u\in M$, we have that $q_u(\delta_x)=0$, where
$$q_u:\dist{M}\To\dist{M}/\big(I_u\cdot\dist{M}\big)$$ is the
quotient mapping and $I_u$ is an ideal in the algebra
$C^{\infty}(M)$ consisting of the functions equal to $0$ at the
point $u$.
\end{lem}
\textbf{Proof.} We have to prove that for any point $u\in M$, there exists
such generalized function $\eta\in\dist{M}$ and a smooth function
$\phi\in I_u$ that: $\delta_x=\phi\cdot\eta$.\\
Consider two cases: $u\neq x$ and $u=x$.
\begin{list}{}{\topsep=10pt\itemsep=10pt}
\item[$\mathbf{u\neq x}:$]
In this case, the function $\phi$, can be any smooth function such
that: $\phi(u)=0$ and $\phi(x)=1$. It is clear that $\phi\in I_u$.
Let us check that the equality, $\delta_x=\phi\cdot\delta_x$ is
true:\\
for any $\psi\in\comp{M}$, we have
$$(\phi\cdot\delta_x)(\psi)=\delta_x(\phi\cdot\psi)=(\phi\cdot\psi)(x)=\phi(x)\cdot\psi(x)=\psi(x)=\delta_x(\psi)$$%
therefore, we obtain that when $u\neq x$, $\delta_x\in I_u$, which
is equivalent to $q_u(\delta_x)=0$ in the quotient module
$\dist{M}/(I_u\cdot\dist{M})$.
\item[$\mathbf{u=x}:$]
In this case, we have that: $\delta_x=-\phi\cdot\delta_x'$, where
the function $\phi$ is any smooth function such that $\phi(x)=0$
and $\phi'(x)=1$:\\
for any $\psi\in\comp{M}$, we have the following
$$
\begin{array}{l}
\\{}
(-\phi\cdot\delta_x')(\psi)=-\delta_x'(\phi\cdot\psi)=\\
\\
=(\phi\cdot\psi)'(x)=\phi'(x)\cdot\psi(x)+\phi(x)\cdot\psi'(x)=\psi(x)=\delta_x(\psi)
\\{}
\end{array}
$$
therefore, the Dirac function $\delta_x$ is the element of the
submodule $I_x~\cdot~\dist{M}$, or equivalently $q_x(\delta_x)=0$,
which finishes the proof.
\end{list}
$\Box$

It follows from this lemma that, for any $x\in M$, the generalized
function $\delta_x$ is such that $q_u(\delta_x)=0$, but the
functional $\delta_x$ obviously is not equal to zero. This fact
implies that the $C^{\infty}(M)$-module \dist{M} \textbf{is not
geometric}.
\subsection{Poisson Bracket on Generalized Functions and Generalized Casimir Functions}
Let $M$ be a Poisson manifold. The action of the vector fields on
the manifold $M$ on the elements of the space \dist{M}, defines
the Poisson bracket of a smooth function $f$ and a \genf{} $\Phi$
as:
$$\poiss{f}{\Phi}=X_f(\Phi)$$
where $X_f$ is the Hamiltonian vector field corresponding to the
function $f$. The following more detailed definition can be used
too: for any $\Phi\in\dist{M}$, $f\in C^{\infty}(M)$ and
$\psi\in\comp{M}$ let
\begin{equation}\label{FunctionDistributionBracket.Formula}
\begin{array}{l}
\\{}
\scalar{\poiss{f}{\Phi}}{\psi}=\scalar{X_f(\Phi)}{\psi}=\\
\\
=-\scalar{\Phi}{X_f(\psi)}=\scalar{\Phi}{\poiss{\psi}{f}}
\\{}
\end{array}
\end{equation}
Hence, it can be stated that we have a connection for the pair
$\big(\,C^{\infty}\,,\,\dist{M}\big)$, that is: for any $f,g\in
C^{\infty}(M)$ and $\Phi\in\dist{M}$
$$
\poiss{f}{g\cdot\Phi}=\poiss{f}{g}\cdot\Phi+g\cdot\poiss{f}{\Phi}
$$
which easily follows from the corresponding property for the
action of vector fields on \genf.

If we define $\poiss{\Phi}{f}$ as $-\poiss{f}{\Phi}$ and consider
the following operator
$\op{\Phi}=\poiss{\Phi}{\cdot}:C^{\infty}(M)\To\dist{M}$, for any
fixed $\Phi\in\dist{M}$, it turns out that $\op{\Phi}$ is a
first-order differential operator, with property
$$
\op{\Phi}(\phi\psi)=\phi\op{\Phi}(\psi)+\psi\op{\Phi}(\phi)
$$
The above equality easily follows from
$$
\{\phi\psi\,,\,\cdot\,\}=\phi\{\psi\,,\,\cdot\,\}+\psi\{\phi\,,\,\cdot\,\}%
$$
Now, when we have already defined the Poisson bracket of a \genf
and a smooth function on a Poisson manifold $M$, it can be stated
that, if the Poisson structure on the manifold $M$ is such that it
is singular but its center coincides with the set of constant
functions on $M$, then it has non-constant center in the space of
\genf, on the manifold $M$. That is, there can be found such \genf
$\Phi\in\dist{M}$, that $\poiss{\Phi}{\psi}=0$, for every
$\psi\in\smooth{(M).}$

In the situation described in the Example~\ref{Example1}, the
distributions commuting with every smooth function are the Dirac
functionals $\delta_a$ for $a\in\varphi^{-1}(0)$. In this case,
for any $f\in\smooth{(M)}$ and $\psi\in\comp{M}$, we have the
following
$$
\begin{array}{l}
\scalar{\poiss{\delta_a}{f}_1}{\psi}=\scalar{\delta_a}{\poiss{f}{\psi}_1}=\\
\\
=\scalar{\delta_a}{\varphi\poiss{f}{\psi}}=\varphi(a)\poiss{f}{\psi}(a)=0
\end{array}
$$
Now, we shall describe some general construction to build the
distributions ''commuting`` with all smooth functions on the
Poisson manifold $M$.

Let $N$ be a symplectic manifold, i.e., the involutive bivector
field correspondent to the Poisson structure on this manifold is
non-degenerated. It is the same that the Poisson bracket is
defined by some symplectic form $\omega$ as
$\poiss{f}{g}=\omega(X_f,X_g)$, for $f,g\in\smooth{(M),}$ where
$X_f$ and $X_g$ are the Hamiltonian vector fields corresponding to
the functions $f$ and $g$: $\ud f=-i_{X_f}\omega$, $\ud
g=-i_{X_g}\omega$.

Let us recall the following formula for the Poisson bracket
\begin{equation}\label{PoissonBracketNew.Formula}
\poiss{f}{g}\cdot\omega^n=n\cdot\ud g\wedge\ud f\wedge\omega^{n-1}%
\end{equation}
where $n=\frac{1}{2}\cdot\textbf{dim}(M)$ and
$\omega^n=\underbrace{\omega\wedge\cdots\wedge\omega}_{n-\textrm{times}}$\
.\\
This formula is the result of the following
$$
\begin{array}{l}
\\{}
\der[g]\wedge\omega^n=0\then\\
\\
\then
i_{X_f}\big(\der[g]\wedge\omega^n\big)=\poiss{f}{g}\cdot\omega^n-\der{g}\wedge
i_{X_f}\big(\omega^n\big)=0\then\\%
\\
\then\poiss{f}{g}\cdot\omega^n=\der{g}\wedge i_{X_f}\big(\omega^n\big)%
\\{}
\end{array}
$$
Let $M$ be a smooth manifold with Poisson structure defined by a
bivector field $p\in V^2(M)$ and $N$ be a symplectic leaf in the
Poisson manifold $M$. That is, the submanifold $N$ is integral for
the distribution defined by the Hamiltonian vector fields and the
restriction of the bivector field $p$, on the leaf $N$ is
non-degenerated. Therefore $p|_N$ corresponds to some symplectic
form on $N$ which we denote under $\omega_N$. Suppose, for
convenience, that the manifold $M$ is compact (which implies that
\smooth{(M)}\,=\,\comp{M}) and the submanifold $N$ is closed ($\cd
N=0$). Consider the following \genf{} on the manifold $M$:
$$
\delta_N:\smooth{(M)}\To\Real\;,\qquad\scalar{\delta_N}{\varphi}=\int\limits_N\varphi|_N\cdot\omega^k%
$$
where $\varphi\in\smooth{(M)}\;,\;k=\frac{1}{2}\cdot\mandim{M}$
and $\varphi|_N$ denotes the restriction of the function $\varphi$
to the submanifold $N$.
\begin{prop}
For any $\varphi\in\smooth{(M)}$, we have that
$\poiss{\delta_N}{\varphi}\,=\,0$
\end{prop}
\textbf{Proof.} By definition of the Poisson bracket of a \genf{} and a
smooth function on a Poisson manifold we have that for any
$\varphi,\psi\in\smooth{(M)}$
$$
\scalar{\poiss{\delta_N}{\varphi}}{\psi}\,=\,\scalar{\delta_N}{\poiss{\varphi}{\psi}}\,=\,\int\limits_N\poiss{\varphi}{\psi}\!\big|_N\cdot\omega_N^k%
$$
Take into consideration the fact that the Hamiltonian vector
fields are tangent to the symplectic leaves, the
formula~\ref{PoissonBracketNew.Formula} and the Stokes formula, we
obtain the following:
$$
\begin{array}{l}
\\{}
\int\limits_N\poiss{\varphi}{\psi}\!\big|_N\cdot\omega_N^k\,=\,\int\limits_N\poiss{\varphi|_N}{\psi|_N}\cdot\omega_N^k\,=\\
\\
=n\cdot\int\limits_N\der{\psi}\wedge\der{\varphi}\wedge\omega_N^{k-1}\,=\,n\cdot\int\limits_{\cd
N}\psi\wedge\der{\varphi}\wedge\omega_N^{k-1}\,=\,0
\\{}
\end{array}
$$
$\Box$

\subsection{The Canonical Comlex of a Poisson Manifold and Generalized Casimir Functions}
For a Poisson manifold $M$, with a bivector field $p\in V^2(M)$
such that $\poiss{f}{g}=i_p(\der{f}\wedge\der{g})$, Koszul
introduced the differential
$$
\delta=i_p\circ\der-\der\circ i_p:\Omega^n(M)\To\Omega^{n-1}(M)
$$
(see Section~\ref{Star} for the noncommutative foundation).

Let us enumerate some properties of the operator $\delta$
(see\cite{brylinski}).

The following expression, reveals the relation between the
operator $\delta$ and the boundary operator for the Lie algebra
homologies:
\begin{equation}\label{DeltaOperator.Formula}
\begin{array}{l}
\\{}
\delta\big(f_0\,\der{f_1}\wedge\cdots\wedge\der{f_k}\big)=\\
\\
=\sum\limits_{1\leq i\leq
k}(-1)^{i+1}\poiss{f_0}{f_i}\,\der{f_1}\wedge\cdots\wedge\widehat{\der{f_i}}\wedge\cdots\wedge\der{f_k}+\\
\\
+\sum\limits_{1\leq i<j\leq
k}(-1)^{i+j}f_0\,\der{\poiss{f_i}{f_j}}\wedge\cdots\wedge\widehat{\der{f_i}}\wedge\cdots\wedge\widehat{\der{f_j}}\wedge\cdots\wedge\der{f_k}%
\\{}
\end{array}
\end{equation}
If we denote the Chevalley-Eilenberg complex of the Lie algebra
$L=\smooth{(M),}$ by $C(\,L\,,\,L\,)$ it can be stated that
$C_k(\,L\,,\,L\,)=L\oplus(\wedge^kL)$ and the differential
$\delta:C_k(\,L\,,\,L\,)\To C_{k-1}(\,L\,,\,L\,)$ is given by the
formula:
$$
\begin{array}{l}
\\{}
\delta\big(f_0\otimes\big(f_1\wedge\cdots\wedge f_k\big)\big)=\\
\\
=\sum\limits_{1\leq i\leq
k}(-1)^{i+1}\poiss{f_0}{f_i}\otimes\big(f_1\wedge\cdots\wedge\widehat{f_i}\wedge\cdots\wedge f_k\big)+\\
\\
+\sum\limits_{1\leq i<j\leq
k}(-1)^{i+j}f_0\otimes\big(\poiss{f_i}{f_j}\wedge\cdots\wedge\widehat{f_i}\wedge\cdots\wedge\widehat{f_j}\wedge\cdots\wedge
f_k\big)%
\\{}
\end{array}
$$
The series of linear mappings:
$\pi_n:C_n(\,L\,,\,L\,)\To\Omega^n(M)$, defined as
$$
\pi_n\big(f_0\otimes\big(f_1\wedge\cdots\wedge f_k\big)\big)=f_0\,\der{f_1}\wedge\cdots\wedge\der{f_k}%
$$
is a homomorphism of the differential complexes
$\big(\;C(\,L\,,\,L\,)\;,\;\delta\;\big)$ and
$\big(\;\Omega(M)\;,\;\delta\;\big)$. That is, for any
$n=0,\ldots,\infty$, we have
$\pi_n\circ\delta=\delta\circ\pi_{n+1}$.

The differential complex
$$
\cdots\;\To\;\Omega^{n+1}(M)\;\stackrel{\delta}{\To}\;\Omega^n(M)\;\To\;\cdots
$$
is called the \textbf{canonical complex} of the Poisson manifold
$M$. The homology of this complex is denoted by
$H^{\textrm{can}}(M)$ and called the \textbf{canonical homology}
of the Poisson manifold
$\big(\,M\,,\,\poiss{\cdot}{\cdot}\,\big)$.

Using the formula~\ref{DeltaOperator.Formula} it is easy to show
that: $\der\circ\delta+\delta\circ\der=0$.

If $\alpha$ is a closed differential form on the manifold $M$,
from the Koszul's definition of of the operator $\delta$ (see
Formula~\ref{PoissonBoundaryOperator.Formula}) immediately follows
that the form $\delta(\alpha)$ is an exact form.

The bivector field $p\in V^2(M)$, corresponding to the Poisson
structure on the manifold $M$, defines a bilinear pairing for any
$k=1,\ldots,\infty$
$$
\wedge^k(\hat{p}):\wedge^k\big(\ctang{M}\big)\otimes\wedge^k\big(\ctang{M}\big)\To\smooth{M}
$$
by the formula:
$$\wedge^k(\hat{p})\big(\alpha\otimes\beta\big)=\big(\alpha\wedge\beta\big)\big(\wedge^kp\big)$$
This mapping is $(-1)^k$-symmetric.\\
As in the case of a Riemannian manifold, it can be defined the
$\star$ operator in the case of symplectic manifold
$$
\star:\Omega^k(M)\To\Omega^{2n-k}(M)
$$
by the formula
$$\beta\wedge(\star(\alpha))=\wedge^k(\hat{p})(\beta,\alpha)\cdot\textbf{vol}$$
where $2n=\textbf{dim}{M};\quad\alpha,\beta\in\Omega^k(M)\quad$
and $\quad\textbf{vol}=\frac{1}{m!}\cdot\omega^m$.
\begin{rem}
The noncommutative definition of the $\star$ operator is given in
the section~\ref{Star}.
\end{rem}
The operator $\star$ is involutive: $\star\circ\star=\textrm{Id}$.
\begin{thm}[see~\cite{brylinski}]
The relation $\delta=(-1)^k\star\circ\der\circ\star$ holds on
$\Omega^k(M)$ for any integer $k\geq 0$.
\end{thm}
\begin{cor}[see~\cite{brylinski}]
For a symplectic manifold, the operator $\star$ defines an
isomorphism of the canonical homology
$H_{\bullet}^{\textrm{can}}(M)$ with the de Rham cohomology
$H^{2m-{\bullet}}(M)$, where $m$ is the dimension of this
manifold.
\end{cor}

Let \dist[0]{M} be the subspace of the space \dist{M} consisting
of the \genf{} commuting with every smooth function on the
manifold $M$;\\
$H_0(\,M\,,\,\delta\,)$ be the space of the $0$-dimensional
homologies of the canonical complex of the Poisson manifold $M$;\\
$H_0^{\textrm{can}}(\,M\,)^*$ be the space of the linear
functionals on the space $H_0^\textrm{can}(\,M\,)$.
\begin{prop}
The spaces \dist[0]{M} and $H_0^{\textrm{can}}(\,M\,)^*$ are
isomorphic.
\end{prop}
\textbf{Proof.} As it follows from the definition of the Poisson bracket of
a distribution and a smooth function, the space $\dist[0]{M}$ can
be defined as
$$
\dist[0]{M}=\{\,\Phi\in\dist{M}\quad|\quad\scalar{\Phi}{\poiss{f}{g}}=0\quad\forall
f,\,\forall g\in\smooth{M}\,\}
$$
In other words:
$\dist[0]{M}=\poiss{\smooth{M}}{\smooth{M}}^{\perp}$, where
$\poiss{\smooth{M}}{\smooth{M}}$ denotes the space of the sums of
the type
$$
\sum\limits_{i,j}\;\poiss{\varphi_i}{\psi_j},\quad\varphi,\psi\in\smooth{M}
$$
As it follows from the formula~\ref{DeltaOperator.Formula} for the
canonical coboundary operator $$\delta:\Omega(M)\To\Omega(M)$$ its
action on the form
$\alpha=\sum\limits_i\;\varphi\der{\psi}\in\Omega^1(M)$ is
$\delta(\alpha)=\sum\limits_i\poiss{\varphi}{\psi}$.\\
Therefore:
$$\delta\big(\Omega^1(M)\big)=\poiss{\smooth{M}}{\smooth{M}}$$
But, by definition, we have that
$$
H_0\big(M,\delta\big)=\smooth{M}\big/\delta\big(\Omega^1(M)\big)
$$
hence, we obtain that
$\delta\big(\Omega^1(M)\big)^{\perp}=H_0\big(\,M\,,\,\delta\,\big)^*$.
$\Box$

\begin{cor}\label{Corollary1.Corollary}
For a compact symplectic manifold $M$ (i.e., the bivector field,
corresponding to the Poisson bracket is non-degenerated), the
space \dist[0]{M} is one-dimensional and the functional
$\scalar{\delta_{\omega}}{\varphi}=\int\limits_M\varphi\cdot\omega^n$,
where $\varphi\in\smooth{(M)}$, $\omega$ is a symplectic form and
$n=\frac{1}{2}\cdot\textbf{dim}(M)$, is a basis of the space
\dist[0]{M}.
\end{cor}
\textbf{Proof.} As $M$ is a symplectic manifold, then the mapping
$$
\star\,:\,H_0^{\textrm{can}}(M)\;\To\;H^{2m}(M)
$$
where $H^{2m}(M)$ is the $2m$-dimensional de Rham cohomology space
of the manifold $M$, is isomorphism. As the manifold $M$ is
symplectic, it is an oriented manifold; that is:
$H^{2m}(M)\iso\Real$.
$\Box$

Let $N$ be a symplectic leaf in the Poisson manifold $M$,
and
$$r:\smooth{(M)}\To\smooth{(N)}$$
be the restriction mapping. It is clear that
$\delta_N=r^{*}(\delta _{\omega_N})$, where
$$r^{*}~:~\dist{N}~\To~\dist{M}$$ is the dual mapping, and
$\omega_N$ is the symplectic form on $N$ induced by the
restriction of the bivector field on the submanifold $N$. If the
mapping $r$ is an epimorphism, then
$$
\textbf{Image}(r^{*})~=~(I_N)^{\perp}
$$
where $I_N$ is the ideal of the functions on $M$ vanishing on the
submanifold $N$, and $(I_N)^{\perp}$ is its orthogonal subspace in
the space $\dist{M}$.
\begin{prop}
If a symplectic leaf $N$ in the Poisson manifold $M$ is such that
the restriction mapping $r:\smooth{(M)}\To\smooth{(N)}$ is an
epimorphic, then the space $(I_N)^{\perp}\cap\dist[0]{M}$ is
one-dimensional and the set $\big\{\,\delta_N\,\big\}$ gives its
basis.
\end{prop}
\textbf{Proof.} As the mapping $r$ is a Poisson mapping (i.e., a
homomorphism of the Poisson algebras), we have that:
$r\big(\poiss{\varphi}{\psi}\big)=\poiss{r(\varphi)}{\psi}$, and
therefore:
$(r^*)^{-1}\big((I_N)^{\perp}\cap\dist[0]{M}\big)=\dist[0]{N}$,
which is one-dimensional according to the
corollary~\ref{Corollary1.Corollary}.
$\Box$

\subsection{Poisson Ideal and Reduction of Poisson Algebra.}
Let $A$ be an associative Poisson algebra. That is: $A$ is an
associative real or complex algebra and a Lie algebra with a
commutator $\{\,\cdot\,,\,\cdot\,\}:A\times A\To A$, such that:
$$
\{\,a\,,\,b\cdot c\}\;=\;b\cdot\{\,a\,,\,c\}\;+\;\{\,a\,,\,c\}\cdot b%
$$
for all $a,b,c\in A$.
\begin{defn}
A subset $I\subset A$ is called a \emph{Poisson ideal} if $I$ is a
two-sided ideal under the multiplication operation in the
associative algebra $A$ and is a Lie algebra ideal in the Lie
algebra $A$:
$$
(\,\forall\,x\in I,\textrm{ and }\forall\,a\in A)\;\then\;(x\cdot
a\in I,\;a\cdot x\in I,\;\{\,x\,,\,a\,\}\in I)
$$
\end{defn}
If $I$ is a Poisson ideal in the Poisson algebra $A$, then the
quotient space $A/I$ is a Poisson algebra too, and the canonical
projection mapping $q:A\To A/I$ is a homomorphism of Poisson
algebras: $q\{a,b\}=\{q(a),q(b)\}$, for all $a,b\in A$.

If a Poisson ideal $I\subset A$ is such that the quotient algebra
$A/I$ is a submanifold algebra, then the algebra $A/I$ can be
considered a noncommutative analogue of a Poisson submanifold of a
Poisson manifold. In this case, the ideal $I$ will be called the
\emph{Poisson submanifold ideal}.
\begin{defn}
A Poisson structure on a Poisson algebra $A$ is said to be
\emph{non-degenerated}, iff the Poisson algebra $A$ does not
contain any Poisson submanifold ideal besides $\{0\}$ and $A$
itself.
\end{defn}
A Poisson submanifold ideal $I$ in a Poisson algebra $A$, will be
said to be \emph{maximal} if $I\neq A$ and the Poisson submanifold
ideal $I'$, containing the ideal $I$ is only $A$.

Let $A$ and $B$ are associative algebras and $f:A\To B$ be their
homomorphism. We call the homomorphism $f$ the \emph{submanifold
mapping}, if for any submanifold ideal $I\subset B$, the set
$f^{-1}(I)$ is also a submanifold ideal in the algebra $A$.

First of all, let us recall that, for any ideal $I\subset B$, the
set $f^{-1}(I)$ is also an ideal in the algebra $A$; and then,
from the definition of a noncommutative submanifold, follows that
if the homomorphism $f:A\To B$ is a submanifold mapping then for
any such ideal $I\subset B$, that the following sequence of
homomorphisms
$$
0\;\To\;Der_I(B)_0\;\To\;Der_I(B)\;\To\;Der(B/I)\;\To\;0
$$
is short, the following sequence of homomorphisms
$$
0\;\To\;Der_{f^{-1}(I)}(A)_0\;\To\;Der_{f^{-1}(I)}(A)\;\To\;Der(A/f^{-1}(I))\;\To\;0
$$
is also short.
\begin{lem}
If $I\subset A$ is a submanifold ideal, then the canonical
projection mapping $q:A\To A/I$ is a submanifold mapping.
\end{lem}
\textbf{Proof.} Consider any submanifold ideal $I'\subset A/I$. It is
clear, that the quotient algebra $A/q^{-1}(I')$ is canonically
isomorphic to the quotient algebra $S_I/I'$, where $S_I=A/I$. So,
we have the following two exact sequences
$$
0\;\To\;Der_I(A)_0\;\To\;Der_I(A)\;\stackrel{r_1}{\To}\;Der(A/I)\;\To\;0
$$
and
$$
0\;\To\;Der_{I'}(A/I)_0\;\To\;Der_{I'}(A/I)\;\stackrel{r_2}{\To}\;Der(A/q^{-1}(I))\;\To\;0
$$
In this situation, we have to prove that the mapping
$$
r_3\,:\,Der_{q^{-1}(I')}(A)\;\To\;Der(A/q^{-1}(I))
$$
is an epimorphism.\\
Let us recall that the space $Der_{I'}(A/I)$ is defined as the
space of such derivations of the algebra $A/I$, which carries the
ideal $I'$ to itself. Therefore, the space
$r_1^{-1}(Der_{I'}(A/I))$ is a subspace of $Der(A)$, consisting of
such derivatives of the algebra $A$, which carries the ideals $I$
and $q^{-1}(I')$ to itself. It is clear that
$r_1^{-1}(Der_{I'}(A/I))$ is a subspace of $Der_{q^{-1}(I')}(A)$
and the mapping
$$
r_2\circ r_1\,:\,r_1^{-1}(Der_{I'}(A/I))\;\To\;Der(A/q^{-1}(I))
$$
which is an epimorphism, is equal to the restriction of the
mapping $r_3$ to the subspace $r_1^{-1}(Der_{I'}(A/I))$.
$\Box$

\begin{thm}[The reduction of Poisson algebra]
If $A$ is a Poisson algebra and $I\subset A$ is a maximal Poisson
submanifold ideal, then the quotient algebra $Q=A/I$ is
non-degenerated Poisson algebra.
\end{thm}
\textbf{Proof.} Let $I'\subset A/I$ be any Poisson submanifold ideal. As
the canonical projection mapping $q:A\To A/I$ is a submanifold
mapping and is an epimorphism, we have that $q^{-1}(I')$ is a
Poisson submanifold ideal in the algebra $A$ and is not equal to
$A$. As $I$ is a maximal Poisson subamanifold ideal, we have that
$q^{-1}(I')=I$, which implies that $I'=\{0\}$.
$\Box$

\begin{example}
Let $M$ be a symplectic manifold and $X$ be any submodule of the
module $\vecf{M}$. Let us denote by $C_X^{\infty}(M)$, the
subalgebra of the commutative algebra $C^{\infty}(M)$, consisting
of such smooth functions $f\in C^{\infty}(M)$ that
$u(f)=0,\;\forall\;u\in X$ and by $X^{\perp}$, the submodule of
$\vecf{M}$, consisting of such elements $u\in\vecf{M}$, that
$\omega(u,\,X)=\{0\}$, where $\omega$ is the symplectic form on
the manifold $M$.

If the submodule $X^{\perp}$ is involutive (i.e.,
$\big[X^{\perp}\,,\,X^{\perp}\big]=0$), then the algebra
$C_X^{\infty}(M)$ is a Poisson algebra (i.e., a Poisson subalgebra
of $C^{\infty}(M)$). To check this, consider any two elements
$f,g\in C_X^{\infty}(M)$. We have
$$
\begin{array}{l}
\\{}
(X(f)\;=\;X(g)\;=\;0)\Leftrightarrow(\omega(h(f),X)\;=\;\omega(h(g),X)\;=\;0)\Leftrightarrow\\
\\
\Leftrightarrow(h(f),h(g)\in
X^{\perp})\then([h(f),h(g)]\;=\;h(\{f,g\})\in
X^{\perp})\then\\
\\
\then(\omega(h(\{f,g\}),X)\;=\;\{0\})\then(X(\{f,g\})\;=\;0)\then\\
\\
\then(f,g\in C_X^{\infty}(M))
\\{}
\end{array}
$$
In the case when the submodule $X$ is generated by a symplectic
action of some Lie group $G$, the submodule $X^{\perp}$ is
involutive:
$$
\begin{array}{l}
\\{}
u\in X \then L_u\omega=0\then\der{i_u\omega}=0\then\\
\\
\then x\omega(u,y)-y\omega(u,x)-\omega(u,[x,y])=0
\\{}
\end{array}
$$
which implies that, if $x,y\in X^{\perp}$ then $[x,y]\in
X^{\perp}$. Hence, we obtain that in the case of a symplectic
action of some Lie group, the algebra of invariant functions under
this group, is a Poisson subalgebra of $C^{\infty}(M)$.

A Poisson ideal in the Poisson algebra $C_X^{\infty}(M)$ can be
constructed by using of a function (if such function exists)
$\varphi\in C_X^{\infty}(M)$ such that $X^{\perp}(\varphi)=\{0\}$,
as the ideal generated by the function $\varphi$.
\end{example}
\newpage

\end{document}